\documentclass[11pt]{article}
\usepackage{fancyhdr}

\fancyheadoffset{0cm}

\fancypagestyle{plain}{%
   \fancyhf{}%
   \fancyhead[R]{\thepage}%
}

\usepackage{subfig}
\usepackage{graphicx}
\usepackage[flushleft]{threeparttable}
\usepackage{multirow}
\usepackage{multicol}
\usepackage{hyperref}
\usepackage{bm}
\usepackage[margin=1in]{geometry}
\usepackage{booktabs}
\usepackage{arydshln}
\usepackage{caption}
\usepackage{siunitx}
\usepackage[table]{xcolor}
\usepackage{dblfloatfix,float}
\usepackage{amsmath,amssymb}
\usepackage{bbm}
\usepackage[font=small,labelfont=bf,tableposition=top]{caption}

\usepackage{parskip} 
\DeclareCaptionLabelFormat{andtable}{#1~#2  \&  \tablename~\thetable}

\usepackage{natbib}
 \bibpunct[, ]{(}{)}{,}{a}{}{,}%

\DeclareMathOperator*{\argmax}{argmax} 
\newcommand{\code}[1]{\texttt{#1}}

\newcommand{\bx}{\mathbf{x}}
\newcommand{\by}{\mathbf{y}}

\newcommand{\Cov}{\mathrm{Cov}}

\newcommand\tab[1][1cm]{\hspace*{#1}}

\begin{document}

\title{Multi-Output Gaussian Processes for Multi-Population Longevity Modeling}
\author{Nhan Huynh\thanks{Department of Statistics and Applied Probability, University of California at Santa Barbara} and Mike Ludkovski\thanks{Department of Statistics and Applied Probability, University of California at Santa Barbara, \url{ludkovski@pstat.ucsb.edu}}}

  \maketitle \thispagestyle{empty}

\begin{abstract}
We investigate joint modeling of longevity trends using the spatial statistical framework of Gaussian Process regression. Our analysis is motivated by the Human Mortality Database (HMD) that provides unified raw mortality tables for nearly 40 countries. Yet few stochastic models exist for handling more than two populations at a time. To bridge this gap, we leverage a spatial covariance framework from machine learning that treats populations as distinct levels of a factor covariate, explicitly capturing the cross-population dependence. The proposed multi-output Gaussian Process models straightforwardly scale up to a dozen populations and moreover intrinsically generate coherent joint longevity scenarios. In our numerous case studies we investigate predictive gains from aggregating mortality experience across nations and genders, including by borrowing the most recently available ``foreign'' data. We show that in our approach, information fusion leads to more precise (and statistically more credible) forecasts. We implement our models in \texttt{R}, as well as a Bayesian version in \texttt{Stan} that provides further uncertainty quantification regarding the estimated mortality covariance structure. All examples utilize public HMD datasets.
\end{abstract}

\section{Mortality Models across Multiple Populations}\label{sec:intro}

Mortality data are typically collected by jurisdictional areas, such as countries and states. As a result global mortality experience is summarized in dozens of national and sub-national registries, presenting a major data-analysis challenge. The burgeoning Human Mortality Database \citep{HMD} offers a centralized portal to nearly 40 such datasets, yielding a rich source of cross-national longevity trends.

Significant predictive value can be extracted from joint models of these mortality tables. By aggregating data, one hopes to improve prediction accuracy (through better disentangling of trends vis-a-vis ``noise'') and simultaneously reduce model risk by increasing credibility of the forecasts. Moreover, joint models capture information fusion, which is very valuable since mortality data are released asynchronously. With a joint model one can rely on the newly released data of a related foreign population to update and improve the domestic forecast. Last but not least, joint models are critical for generating forecasts and future scenarios simultaneously across multiple populations. Individual models will tend to be non-coherent, i.e.~include scenarios where the joint mortality trends cross-over or diverge in unrealistic ways.

Yet few models exist for predictive multi-population longevity analysis beyond two populations. The latter case affords the convenient hierarchy of treating one population as the baseline ``index'' and then modeling the ``spread'' between the index and the secondary population. With three or more populations the conceptual meaning of the multiple resulting longevity spreads becomes fuzzy. Moreover, in the commonly adopted Age-Period-Cohort-style models, multiple populations are treated through decomposition into global- and population-specific factors, implying that the number of factors grows linearly in the number of populations. Since each factor (Age, Period, etc) contains dozens of parameters, one quickly ends up with hundreds of parameters to be estimated, creating significant computation and statistical inference  bottlenecks.

In this article we investigate a scalable machine learning approach that simultaneously models all the longevity surfaces within a joint \emph{spatial covariance} framework. We treat populations as a factor covariate, with the respective correlation inferred as part of the model fitting procedure.  Specifically, we employ multi-output Gaussian Process (GP) models, building upon our earlier work in \cite{LUDKOVSKI2016GAUSSIAN} and \cite{HLZ20} on the use of GPs for longevity predictive analytics.  GPs treat age-specific mortality rates as a noisily observed response surface that is learned via the multivariate \emph{kriging} equations. Embedding multiple populations within a multi-output GP naturally captures the borrowing of mortality information and the underlying commonality of mortality patterns. Indeed, a multi-output GP imposes a transparent correlation structure on the co-dependence of mortality rates across populations, disentangling it from the global Age-Year pattern. Moreover, GPs afford a Bayesian perspective, yielding a full uncertainty quantification---including coherent multi-population stochastic scenarios---not just for mortality rates, but also for mortality improvement factors.

The GP paradigm brings a flexible non-parametric spatio-temporal structure that treats tasks of data smoothing (aka in-sample prediction) and forecasting (aka out-of-sample prediction) in an internally-consistent manner. The vast GP ecosystem has become a centerpiece of probabilistic data science and includes a multitude of extensions, from GP GLMs (to address non-Gaussian observation noise) to Kronecker GP (for faster analysis of gridded data). We refer to  \cite{Duvenaud2014,Chu2005,GarridoMerchn2018} for further discussion of GP modeling with categorical covariates. From the machine learning perspective, the respective ideas of multi-task learning and
transfer learning have wide applications \citep{Bonilla2008, Caruana1997,letham2019bayesian}. 

 In the context of mortality modeling, single-population GPs were investigated in  \cite{LUDKOVSKI2016GAUSSIAN} and some preliminary multi-population analysis appeared recently in \cite{HLZ20}. Related spatio-temporal methods were investigated by \cite{christiansen2015differences} to capture the spread between individual log mortality rates and weighted average log-mortality and \cite{debon2010geostatistical}. Another related analysis of the HMD can be found in \cite{CARRACEDO2018} who applied spatio-temporal Markov clustering to detect common patterns of longevity across 26 European countries; see also \cite{antonio2017producing}.  Another way to introduce dependence between populations is through statistical shrinkage within a Bayesian hierarchical model. \cite{raftery2012bayesian} modeled mortality of 160+ countries by first imposing a global hyper-prior over several one-dimensional parameters and then constructing individual Lee-Carter models. A related approach is taken up by \cite{wisniowski2015bayesian}. This framework also permits to inject additional socio-economic or geo-political covariates to capture the varying degree of dependence \citep{kleinow2013mortality,boonen2017modeling}.

The strand of literature addressing  multi-population extensions of the now-classical Lee-Carter stochastic mortality framework is getting longer. The seminal work by \cite{LI2005} extended Lee-Carter to two populations, postulating a decomposition of mortality into population-specific plus global Age and Period factors (for a total of $2L+2$ factors with $L$ populations). More parsimonious versions were proposed by \cite{KLEINOW2015} who considered a common Age effect, and \cite{delwarde2006application} who proposed a common Period effect. \cite{ENCHEV2017} investigated several intermediate cases. Dispensing with country-specific factors allows more interpretability, e.g.~in the \cite{KLEINOW2015} CAE model one may directly compare period effects across countries since these  are scaled with the same age parameters. From the other direction, the model of \cite{LI2005} functionally corresponds to having a single degree of freedom in the evolution of the mortality curve over time. According to \cite{LI2013} at least two Age/Period factors are warranted, and accordingly a multi-factor extension was investigated. Note that in our setup mortality curves are non-parametric (i.e.~as many degrees of freedom as there are data points). To achieve coherent forecasts with  two populations,
 \cite{hyndman2013coherent} considered co-integration; see also the multi-level functional regression approach in \cite{shang2016mortality}. In a similar spirit, \cite{d2016multiple,li2017coherent,guibert2017forecasting}  investigated Vector Autoregressive (VAR) approaches to achieve correlation across the multiple Period factors of the aforementioned \cite{LI2005} framework. Alternatively, \cite{chen2015multi,WANG2015} applied a copula approach to capture the dependence between individual Period factors and \cite{yang2013pricing} considered a Vector Error Correction model.

To sum up, our main contribution is a statistical methodology to build multi-population longevity models through multi-output GP (MOGP). A key innovation is the use of ICM kernels to make more efficient and scalable models and also to achieve dimension reduction. Another innovation is a detailed discussion about how to pool populations to maximize predictive power. The analysis below supercedes the earlier proceedings version in \cite{HLZ20} that concentrated on descriptive investigation of single-population GPs across HMD datasets and primarily focused on 2-population cases with full-rank kernels.

The rest of the paper is organized as follows. 
Section \ref{sec:gp} describes the MOGP model for multiple-population mortality analysis. Section \ref{sec:smape} focuses on how MOGPs can maximize predictive accuracy, while Section \ref{sec:features} shows how MOGP is appropriate for coherent forecasting and capturing commonality of mortality experience.

\subsection{Motivating Multi-Population Models}\label{sec:motivation}
Our primary motivation for developing multi-population models is to improve predictive accuracy. It is generally accepted that there is strong commonality in mortality experiences of different populations, and therefore that there is an opportunity for data fusion to better capture trends and de-noise raw data. Hence, we wish to build a multi-population model to maximize its ``credibility", or equivalently reduce the mis-specification between the true mortality evolution and the fitted model dynamics that arises due to using limited historical data. The latter idea has several complementary aspects that we now enumerate to foreshadow the methods and results below.

First, data fusion is intrinsically linked to data selection: one should only model populations that are actually dependent; including those that are little-correlated is likely to worsen model fit and forecasting performance. Therefore, multi-population modeling is closely linked to identifying dependence patterns and selecting (i.e.~clustering) populations that are most correlated. From a modeling perspective, judicious choice of which populations to aggregate is critical to keeping models tractable and computationally lean---it is beyond the reach of nearly all methods to directly handle the full HMD with 75+ datasets.

Second, data fusion is also important for mitigating model risk, i.e.~for fitting the best model. Therefore, successful data fusion is expected to manifest itself in reduced parameter uncertainty. For GPs, this translates into tighter hyperparameter and latent-surface posteriors, affording a transparent visualization of higher model stability and higher confidence into the predictive forecasts. We emphasize the latter Bayesian concepts and assess our predictive accuracy not simply through the predictive mean (via a mean-squared-error related metric) but also through scoring rules, such as CRPS that are based on the quality of the full predictive distribution.

Third, for HMD data in particular, a key problem is to obtain highest-quality contemporary forecasts, i.e.~to assess the present-dat mortality in a given ``domestic" population. Because the database is updated very frequently and asynchronously across different countries, at any given timepoint the database is not rectangular but ``notched", i.e. it ends at different years. For instance, as of February 2020 some countries already have 2018 data added, most have data up to 2017, and a few are still lagging and only have data up to 2016. Using such input data to build the best-feasible prediction of 2019 mortality in say UK is possibly the most common use of HMD, but presents challenges relative to the ``classical" time-series models. As we show, GPs are both easily adaptable and internally consistent for  this task.

To conclude this section, we reiterate the big picture perspective that underlies the premise of multi-population analysis: because populations \emph{are} similar---both in their static structure and in their dynamic evolution---one should leverage this similarity to improve predictive analysis. But this must be done in a smart way, keeping in mind computational and statistical considerations and the specific predictive tasks envisioned. We believe that for mortality analysis, a multi-output machine learning framework checks off all these boxes and offers a tractable and scalable way to jointly analyze a collection of 2-10 populations.

\textbf{Data Source:} We work with mortality data from the \href{https://www.mortality.org}{Human Mortality Database} (HMD) \citep{HMD} which provides the aggregated mortality statistics at the national levels for 40 developed countries across the globe. The HMD applies the same consistent set of procedures \citep{HMD2015} on each population and presently focuses on developed economies where death registrations and census data are available and reliable. For our analysis we rely on one-year age groups, concentrating on Ages 50--84 (retirement ages most relevant for predictive actuarial analysis) for both genders and calendar Years 1990--2016. In the models below we consider various subsets of the following 16 European datasets: Austria (AUT), Belgium (BEL), Czech Republic (CZE), Denmark (DEN), Estonia (EST), France (FRA), Germany (GER), Hungary (HUN), Latvia (LAT), Lithuania (LTU), Netherlands (NED), Poland (POL), Spain (ESP), Sweden (SWE), Switzerland (SUI) and United Kingdom (UK or GBR).

The dataset is organized as a large table. The $n$th observation for the $l$th country contains (i) Age and Year as a pair of independent variables, $(x_{ag}^n,x_{yr}^n)$, and (ii) the logarithm of the observed mortality rate, 
\begin{equation}
    y^n = \log\bigg[\frac{\text{Death counts at $(x_{ag}^n,x_{yr}^n)$}}{\text{Exposed-to-risk counts at $(x_{ag}^n,x_{yr}^n)$}}\bigg]=\log\bigg[\frac{D^n}{E^n}\bigg].
\end{equation}
We denote by $\mathcal{D}_l =\{(x^n,y^n)\}_{n=1}^N$ the dataset for the $l$th country.

\section{Multi-Output Gaussian Process Models}\label{sec:gp}

\subsection{Gaussian Process Regression for Mortality Tables}
\label{GPmethod}

A Gaussian process (GP) is an infinite collection of random variables, any finite number of which follows a multivariate normal (MVN) distribution. As such, a GP $f \sim GP( m, C)$ is characterized by its mean function $m(x)$ and its covariance structure $C(x,x')$. This means that for any vector $\mathbf{x}=(x^1,\ldots,x^n)$ of $n$ inputs:
\begin{equation*}
    f(x^1),\ldots,f(x^n) \sim \mathcal{N}\big(\mathbf{m(x)},\mathbf{C(x,x)}\big)
\end{equation*}
where $\mathbf{m(x)}=\mathbb{E}[f(\mathbf{x})]$ is the mean vector of size $n$ and $\mathbf{C(x,x)}$ is the $n$ by $n$ covariance matrix, $C(x,x') := \mathbb{E}[(f({x})-{m(x)})(f({x'})-{m(x')})]$.

In a GP regression setup, the latent $f$ links the observations or output vector $\mathbf{y}=(y^1,\ldots,y^n)$ to the input vector $\mathbf{x}$ via:
\begin{equation}
\label{eqn:regression}
    y^i = f(x^i) + \epsilon^i,
\end{equation}
where $\epsilon^i$ is the error term to reflect that we observe only a noisy sample of $f(x^i)$. In our context, $x^i$ are the individual cells in a mortality table (so indexed by Age, Year, etc.), $y^i$ are observed raw log mortality rates, and $f(x^i)$ is the \emph{true} mortality rate that would materialize in the absence of any random shocks. We assume that observation noise is Gaussian: $\epsilon^i \sim \mathcal{N}(0,\sigma^2)$ or $\epsilon=(\epsilon^1,...,\epsilon^n) \sim \mathcal{N}(\mathbf{0},\mathbf{\Sigma}=\text{diag}(\sigma^2))$. It follows that $\Cov(y^i,y^j) = \Cov(f(x^i),f(x^j)) + \sigma^2\delta(x^i,x^j)$ and therefore $\by \sim \mathcal{N}( \mathbf{m(x)}, \mathbf{C(x,x) + \Sigma})$ where $\delta(x^i,x^j)$ is the Kronecker delta.

GP regression works by applying the Bayesian formalism of assigning a prior distribution to $f \sim GP( m, C)$ and using \emph{MVN conditioning} relative to a dataset $\mathcal{D}=(\bx,\by)$ to infer the posterior distribution of $f$. The Gaussian structure of the prior and the Gaussian structure of \eqref{eqn:regression} together with Bayes' rule yield a Gaussian posterior $p(\mathbf{f|\mathcal{D}})  \propto \text{ }p(\mathbf{f})p(\mathbf{y|x},\Theta)$:
\begin{equation*}
\begin{aligned}
    \text{Posterior distribution} &\ = \frac{\text{Prior distribution x Likelihood function}}{\text{Marginal distribution}}. 
\end{aligned}
\end{equation*}
The principal objective is to draw prediction about $\mathbf{f_*} \equiv f(\mathbf{x_*})$ or future observations $\mathbf{y_*} \equiv Y(\mathbf{x_*})$ at new inputs $\mathbf{x_*}$. By construction,
$\mathbf{y}$ and $\mathbf{y_*}$ follow a joint MVN distribution:
\begin{equation*}
\begin{bmatrix}
    \mathbf{y} \\
    \mathbf{y_*}
 \end{bmatrix}
 \sim \mathcal{N}\Bigg(
 \begin{bmatrix}
     \mathbf{m} \\
     \mathbf{m_{*}}
 \end{bmatrix},
 \begin{bmatrix}
     \mathbf{C+\Sigma} & \mathbf{C(x,x_*)} \\
     \mathbf{C(x,x_*)}^T & \mathbf{C_{**}+\Sigma_{**}}
 \end{bmatrix}
 \Bigg)
\end{equation*}
where $\mathbf{C(x,x_*)}$ is the covariance matrix between training inputs $\bx$ and test inputs $\bx_*$, $\mathbf{C}_{**}$ is the covariance matrix of $\bx_*$, $\mathbf{\Sigma_{**}}=\text{diag}(\sigma^2)$ is the noise variance matrix of the test inputs $\bx_*$,
and $\mathbf{m_{*}} = m(\mathbf{x_*})$. 
The MVN formulas then imply that
\begin{align}\notag
    p(\mathbf{y_*|y}) & \sim \mathcal{N}(\mathbf{m_*(x_*),C_*(x_*,x_*)}) \qquad \text{where} \\ \label{eq:mean1}
       \mathbb{E}[\mathbf{y_*|\mathcal{D}}] = \mathbf{m_*(x_*)} &\ = \mathbf{m+C(x,x_*)}^T[\mathbf{C+\Sigma}]^{-1}(\mathbf{y-m}); \\ \label{eq:var1}
    \mathbb{V}ar(\mathbf{y_*}|\mathcal{D}) = \mathbf{C_*(x_*,x_*)} &\ = \mathbf{C_{**}+\Sigma_{**}}-\mathbf{C(x,x_*)}^T[\mathbf{C+\Sigma]}^{-1}\mathbf{C(x_*,x)}.
\end{align}
Note that the posterior variance $\mathbf{C_*(x_*,x_*)}$ is equal to the prior variance $\mathbf{C_{**}+\Sigma_{**}}$ minus a positive term which reflects the information gained (relative to the prior) from the training data. Furthermore, \eqref{eq:mean1}-\eqref{eq:var1} are valid for \emph{any} $\bx_*$, i.e.~both for in-sample smoothing or for out-of-sample extrapolation.

\textbf{Covariance function.} The kernel $C(x,x')$ captures the correlation between mortality rate at a given Age and Year and mortality rates at other coordinates. For example, we expect the mortality for age 70 in 2010 or $x^i=(70,2010)$, to be more correlated with $x^j=(69,2011)$ than with $x^j=(50,1995)$. In this paper, we employ the squared-exponential kernel:
    \begin{equation}
    \label{eqn:kernel}
        \tilde{C}(x^i,x^j)= \eta^2\mathrm{exp}\bigg[-\frac{(x^i_{ag}-x^j_{ag})^2}{2\theta_{ag}^2}-\frac{(x^i_{yr}-x^j_{yr})^2}{2\theta_{yr}^2}\bigg].
    \end{equation}
   When $x^i \approx x^j$, the covariance reaches its maximum value $\tilde{C}( x^i,x^j) \le \eta^2$; when $x^i$ and $x^j$ are far apart, $\tilde{C}( x^i,x^j) \approx 0$. This feature of expressing the dependence structure through a spatial perspective is central to GPs and is controlled by the  hyperparameters $\theta_{ag}$ and $\theta_{yr}$ in \eqref{eqn:kernel} that are called characteristic length-scales. The lengthscales determine how much influence an observation has on others in the Age and Year dimensions, respectively. Note that $\theta_{ag}$ ---lengthscale for Age---and $\theta_{yr}$ ---lengthscale for Year---are not comparable. An important aspect that influences the goodness-of-fit of a GP model is its spatial smoothness. The squared exponential covariance kernel \eqref{eqn:kernel} makes the mortality curves infinitely differentiable in both Age and Year dimensions (note that the GP is defined over $x \in \mathbb{R}_+^2$ and so provides a continuous interpolation of the observed data gridded by year). This will be exploited below for computing mortality improvement factors. Moreover, the lengthscales $\theta$ affect the qualitative nature of the fitted $m_*(\cdot)$. When lengthscales are too large, the fitted curves are over-smoothed and the influence of individual data points attenuates \citep{Rasmussen2005}. As a result, there is less flexibility in $m_*(\cdot)$; to compensate, the estimated observation noise is increased and the model under-fits.  In contrast, too small lengthscales indicate over-fitting of the spatial dependence, generating high-frequency oscillations in the fitted $m_*(\cdot)$ and low observation noise $\sigma^2$.

\subsection{Multi-output GP kernels}
\label{jointModel}

The idea of commonality in mortality experiences is equivalent to the existence of global longevity trends.
 In the context of a spatio-temporal model, it implies an underlying \emph{shared} covariance structure. This can be easily verified visually or statistically, see \cite{HLZ20} for comparison of 10+ European countries available in HMD where we observe both similar fitted mortality dynamics (e.g.~similar mortality improvement curves over time) and similar estimated GP covariance hyperparameters.

Let $L$ be the number of different populations considered. To jointly model the $L$ different outputs, $\{f_l\}_{1\leq l \leq L}$ we correlate them using the framework of multi-output Gaussian Processes (MOGP) which was introduced in geostatistics under the name of multivariate  kriging or co-kriging \citep{Chiles.etal1999, VERHOEF1998275}. The aim of co-kriging is to estimate the under-sampled variables using spatial correlation with other sampled variables.

 The vector-valued latent response over the Age-Year input space is defined as:
$$\mathbf{f(x)}=(f_1(x^1),\ldots,f_1(x^N),\ldots,f_L(x^1),\ldots,f_L(x^N))=(f_1(\mathbf{x}),\ldots,f_L(\mathbf{x}))$$ where the functions $\{f_l\}_{l=1}^L$ are the log-mortality surface for the corresponding population $l$. Similar to single-output GP (SOGP), MOGP assumes that the vector-valued $\mathbf{f}$ follows a GP:
$$\mathbf{f} \sim GP(\mathbf{m,C})$$
where $\mathbf{m} \in \mathbb{R}^{LN \times 1}$ is the mean vector whose elements $\{m_l(\mathbf{x})\}_{l=1}^L$ are the mean functions of each output, and $\mathbf{C} \in \mathbb{R}^{LN \times LN}$ is the fused covariance matrix.

Populations are treated as categorical input, encoded via $L$ additional input dimensions with 0/1 encoding.
Thus, the input vector for  the $n$th observation in the joint model is $x^n=(x^n_{ag},x^n_{yr},x^n_{pop,1},\ldots,x^n_{pop,L})$, where $x^n_{pop,l}=\mathbbm{1}_{\{\text{population}=l\}}$ are indicators, set to 1 if and only if the $n$th observation is from population $l$.
We denote by $N_l$ the number of Age-Year inputs for population $l$. If all $L$ populations have the same set of inputs and $N_1=\ldots=N_L=N$, the dataset is said to be \textit{isotropic}.

To construct the fused $\mathbf{C}$, one approach is to take the product between a kernel for the Age-Year covariates $x_{ag}, x_{yr}$ and a kernel for the categorical ones \citep{Qian2008, Roustant2018}. Let:
\begin{align}
    \Gamma_{(l_1,i),(l_2,j)} &\ = \mathrm{exp}\bigg[-\theta_{l_1,l_2}\delta_{l_1,l_2}^{ij}\bigg] \tab \text{where }\quad l_1, l_2 \in \{1,
    \ldots,L\}, \label{discreteKern}
\end{align}
with
$$\delta_{l_1,l_2}^{ij} =
    \begin{cases}
        1 & \text{$i$th and $j$th observation come from populations $l_1$ and $l_2$;}  \\
        0 & \text{otherwise}.
    \end{cases}$$
Note that $\delta_{l_1,l_2}^{ij} = 1_{\{ x^i_{l_1} \neq x^j_{l_1}\}} \cdot 1_{\{ x^i_{l_2} \neq x^j_{l_2} \} }$ is symmetric in $i$ and $j$.

Then, the covariance between input rows $x^i$ and $x^j$ is set as follows:
\begin{align}
    C(x^i,x^j) &\ := \eta^2 \mathrm{exp}\Bigg[-\frac{(x^i_{ag}-x^j_{ag})^2}{2\theta_{ag}^2}-\frac{(x^i_{yr}-x^j_{yr})^2}{2\theta_{yr}^2}\Bigg]\prod_{\{l_1,l_2\} }\mathrm{exp}\Bigg[-\theta_{l_1,l_2}\delta_{l_1,l_2}^{ij}\Bigg] \label{eqn:kernelMulti}\\
    &\ = \begin{cases}
      \tilde{C}_{i,j} & \text{if observations are from the same population}; \\
      \tilde{C}_{i,j} \Gamma_{(l_1,i),(l_2,j)}  & \text{if observations are from populations } l_1, l_2,  \notag
    \end{cases}
\end{align}

When observations are from the same country, the covariance between the $i$th and $j$th observation is the same as in a SOGP model, cf.~Equation \eqref{eqn:kernel}. 
 Intuitively, $\Gamma_{l_1,l_2} <1$ discounts the covariance when observations are from different populations and is driven
by the hyperparameter $\theta_{l_1,l_2}$: large value of $\theta_{l_1,l_2}$ implies low correlation $r_{l_1, l_2} := \mathrm{exp}\big(-\theta_{l_1,l_2}\big)$ between the two populations.

Two important assumptions are made in Equation \eqref{eqn:kernelMulti}. First, there is separability \citep{alvarez2011kernels} between the cross-population covariance and the covariance over the Age-Year inputs. Second, observations across $L$ populations share the same spatial covariance kernel. This assumption is useful to examine the commonality in the mortality across populations via the lengthscales in Age and Year dimensions. It can be thought of as full statistical shrinkage of the individual population covariances towards a common baseline, cf.~Section~\ref{sec:better-hyper} below.

\subsection{Coregionalized Kernels}
Estimating cross-population covariance in \eqref{eqn:kernelMulti} requires fitting $L(L-1)/2$ parameters $\theta_{l_1, l_2}, 1 \le l_1 \le l_2 \le L$ which imposes challenges in both statistical and computational aspects when modelling MOGP with many outputs (e.g.~$L \geq 4$).
An attractive dimension reduction approach to keep the number of correlation parameters low is the intrinsic coregionalization model (ICM) \citep{alvarez2011kernels}. In ICM, each output $f_l$ is assumed be a linear combination of independent latent GPs. Let $u_1(\mathbf{x}),\ldots,u_Q(\mathbf{x})$ be independent latent functions each from a GP prior with covariance kernel $C^{(u)}(\mathbf{x,x'})$. The modeled outputs $f_l$  are linear combinations of these $Q$ latent factors:
\begin{equation}
    f_l(\mathbf{x}) = a_{l,1}u_1(\mathbf{x}) + \ldots + a_{l,Q}u_Q(\mathbf{x}) = \sum_{q=1}^Q a_{l,q}u_q(\mathbf{x}),
\end{equation}
where $a_{l,q}$'s are the factor loadings. Let $\mathbf{a}_q=(a_{1,q},\ldots,a_{L,q})^{T}$ be the vector representing the collection of linear coefficients associated with the q$th$ latent function across the $L$ outputs, so that $ \mathbf{f(x)} = \sum_{q=1}^Q \mathbf{a}_qu_q(\mathbf{x})$. It follows that the covariance for $\mathbf{f(x)}$ is:
\begin{align}
    \mathbf{C}(x,x') = \Cov \left(\mathbf{f(x)},\mathbf{f(x')} \right) & = \Cov\bigg(\sum_{q=1}^Q \mathbf{a}_qu_q(\mathbf{x}),\sum_{q=1}^Q \mathbf{a}_qu_q(\mathbf{x'})\bigg)
    \notag\\
    & = \bigg(\sum_{q=1}^Q \mathbf{a}_q\mathbf{a}_q^T\bigg) \otimes \Cov \left(u_q(\mathbf{x}),u_q(\mathbf{x'}) \right) \notag \\
    & = AA^T \otimes C^{(u)}(\mathbf{x,x'}) \label{eqn:ICM_A}
\end{align}
where matrix $A = (\mathbf{a}_1,\ldots,\mathbf{a}_Q)$ and $\otimes$ is the Kronecker matrix product. Re-parametrizing by $B:=AA^T$, \eqref{eqn:ICM_A} can be expressed as the Kronecker product between the cross-population covariance $B \in \mathbb{R}^{L \times L}$ and the covariance over the Age-Year inputs $\mathbf{C}^{(u)} \in \mathbb{R}^{N \times N}$:
\begin{equation}
    \mathbf{C}=\Cov(\mathbf{f(x),f(x')}) = B \otimes \mathbf{C}^{(u)}.
    \label{eqn:ICM_B}
\end{equation}

The \textit{coregionalization matrix} $B$ has rank $Q$. Under ICM, the number of hyperparameters in the cross-population covariance is reduced from $L(L-1)/2$ to $Q \times L$.
Hence, taking $Q  < L/2$ allows to reduce the hyper-parameter space and alleviate the computational budget compared to the ``full rank'' setup.

ICM is a special form of the linear coregionalization model (LCM) \citep{alvarez2011kernels}. In LCM, the independent latent functions are from different GP priors to capture all possible variability from multiple outputs. Its application is beyond the scope of this paper. Notably the assumption in ICM that all $L$ populations share the common spatial covariance suits our interest in the inference of joint lengthscales in Age and Year dimensions. The computational complexity required in ICM is greatly simplified due to the properties of the Kronecker product. Finally, ICM allows the process variance $\eta^2_l$ to vary by populations,  i.e., the l$th$ element in the diagonal in $B$ is the process variance for population $l$. This further bolsters the flexibility in MOGP to excel in out-of-sample forecasts.

The Kronecker decomposition in \eqref{eqn:ICM_B} is also highly useful to speed up the overall fitting. The most expensive step in building a MOGP is solving for the inverse of the covariance matrix $\mathbf{C}$ that shows up in the log-likelihood function. Indeed, inverting this $LN \times LN$ matrix is computationally expensive even for modest values of $L$ and $N$.  Since the inverse of the Kronecker product is equal to the product of the respective inverses: $\big(B \otimes C^{(u)}\big)^{-1} = B^{-1} \otimes \big(C^{(u)}\big)^{-1}$, the ICM structure reduces the computational complexity from ${\cal O}( (LN)^3 )$ to ${\cal O}( L^3 + N^3 + LN)$. In our experience this translates to a factor of 2-4 speed-up in computational time, allowing us to scale up to 10-12 populations.  We emphasize that the overall MOGP (and ICM) structure is completely agnostic to $L$, so that exactly the same numerical method is applied to handle $L=2$ populations or $L=10$. 

\textbf{Selecting ICM rank $Q$.} Because $Q$ is not one of the hyperparameters to be optimized, we exploit the Bayesian Information Criterion (BIC) to select the value of $Q$ for the most parsimonious models. To illustrate the role of the rank $Q$ of an ICM kernel, Table \ref{tbl:fullvsICM} compares several MOGP models with different $Q=2,3,4$. We consider two case studies, both with $L=8$ but with different constituent populations. First, we note the remarkably faster training time (3-time speedup) in ICM relative to a full rank kernel; notably the speedup is independent of $Q$ and is driven by the Kronecker matrix algebra. Second, we note that most ICM models have better predictive performance (see Sections~\ref{perform-metrics} and \ref{cluster} below for explanation of SMAPE and CRPS) than the kernel from \eqref{eqn:kernelMulti}. Third, BIC criterion suggests that $Q=2$ is the preferred model in both cases. Note that there are only $2L = 16$ hyperparameters in the resulting ICM kernel, almost twice as few as $L(L-1)/2 = 28$ hyperparameters in the respective \eqref{eqn:kernelMulti}.

\textbf{Non-Rectangular Data Sets.} We have discussed the use of ICM for isotropic data. The ICM framework can be extended to deal with {partially heterotropic data} where only a portion of $L$ inputs are available and which arises in HMD due to missing data especially at the most recent years. Let $M'$ be the number of distinct inputs across $L$ populations and $M = N_1+\ldots+N_L$ be the number of observations in training data. We consider the setting that $M' < ML$ so that for some inputs not all $L$ outputs are observed. Define the vector-valued ``complete data" function $\mathbf{f(x)}$, with $\mathbf{f(x)} \in \mathbbm{R}^{LM' \times 1}$. We further introduce $\mathbf{f}^o\mathbf{(x)}$ as the vector-valued function corresponding to the observed outputs, $\mathbf{f}^o\mathbf{(x)} \in \mathbbm{R}^{M \times 1}$. The relation between $\mathbf{f(x)}$ and $\mathbf{f}^o\mathbf{(x)}$ is formulated through the ``communication" matrix $S$,
$\mathbf{f}^o\mathbf{(x)} = S^T\mathbf{f(x)}$,
where $S \in \mathbbm{R}^{LM' \times M}$. The column vectors in $S$ are orthonormal with values of 0 and 1 to eliminate the unobserved outputs, see \cite{Skolidis2011heterotopic}. Applying linear transformation to a MVN vector, we can then identify the distribution of $\mathbf{\mathbf{f}^o\mathbf{(x)}}$ as a GP with covariance: $$\Cov(\mathbf{f}^o\mathbf{(x)},\mathbf{f}^o\mathbf{(x')})=S^T \Cov(\mathbf{f(x)},\mathbf{f(x')})S = S^T(B \otimes K)S,$$ recovering the Kronecker structure

\begin{table}[!t]
\caption{Comparison between full rank and ICM MOGP models. Improvement in SMAPE and CRPS use Hungary as the target population and compare MOGP with respect to SOGP Hungary model. The reported percentages are averages over one-year ahead Hungarian mortality forecasts for Ages 70--84 based on three training sets: 1990--2013 (predict 2014), 1990--2014 (predict 2015), and 1990--2015 (predict 2016) for same Ages 70--84.  Best metrics are in \textit{italics}.}
\label{tbl:fullvsICM}
\centering
\begin{tabular}{lrrrr} \toprule
& Full Rank    & ICM ($Q=2$)  & ICM ($Q=3$)    & ICM ($Q=4$)     \\ \midrule
\# Kernel Hyperparameters & 28  & 16   & 24   & 32            \\ \midrule
\multicolumn{5}{c}{\textbf{Case study I}: AUT,  DEN, EST, GER, HUN, LTU, CZE, POL}   \\ \midrule
Running time (mins)                                                         & 132.75   & 59.51  & 59.41      & 58.86       \\
Improvement in SMAPE (\%)                                                                & $-$2.89  & \textit{10.75}    & $-$1.54  & 1.93      \\
Improvement in CRPS (\%)                                                                 & 2.79     & \textit{16.82}     & 11.09    & 16.63     \\
Total BICs                                                                          & $-$          & \textit{$-$28,073} & $-$27,719 & $-$28,006  \\ \midrule
\multicolumn{5}{c}{\textbf{Case study II}: EST, HUN, LTU, NED, POL,  SWE, SUI, GBR}          \\    \midrule
Running time (mins)                                                         & 176.78     & 59.03    & 58.99      & 58.98       \\
Improvement in SMAPE (\%)                                                           & $-$24.90  &  $-$2.77   & 7.71     & $-$32.76  \\
Improvement in CRPS (\%)                                                            & $-$2.28   & 13.53     & 14.04    & $-$2.95   \\
Total BICs                                                                          & $-$       & \textit{$-$28,890} & $-$28,346  & $-$28,361 \\ \bottomrule
\end{tabular}
\end{table}

\subsection{GP Hyperparameters}
To implement a GP model requires specifying its hyperparameters. Note that actual inference reduces to linear-algebraic formulas in \eqref{eq:GP-mean}-\eqref{eq:GP-var}, and the modeling task is to learn the spatial covariance, namely the mean and kernel functions.

\subsubsection*{Mean function} 
To capture the long-term longevity features, such as higher mortality at higher ages,  we fit a parametric prior mean: $$m(x)=\beta_0+\sum_{j=1}^p \beta_jh_j(x),$$ where $h_j$'s are given basis functions and the $\beta_j$'s are unknown coefficients. The coefficients $\bm{\beta}=(\beta_1,\ldots,\beta_p)^T$ are estimated simultaneously with other hyperparameters. Let $\mathbf{h}(x)=\big(h_1(x), \ldots,h_p(x)\big)$ and $\mathbf{H}=\big(\mathbf{h}(x^1),\ldots,\mathbf{h}(x^N)\big)$, then the posterior mean of $\bm{\beta}$ along with the predicted posterior mean $m_*(x_*)$ and respective variance $s_*^2(x_*)$ for a new input $x_*$ are:
\begin{align}
    \bm{\hat{\beta}} &\ =\mathbf{(H}^T\mathbf{(C+\Sigma)}^{-1}\mathbf{H)}^{-1}\mathbf{H}^T\mathbf{(C+\Sigma)}^{-1}\mathbf{y}; \\ \label{eq:GP-mean}
    m_*(x_*) &\ = \mathbf{h}(x_*)^T\bm{\hat{\beta}}+\mathbf{c}(x_*)^T\mathbf{(C+\Sigma)}^{-1}(\mathbf{y-H}\bm{\hat{\beta}}); \\ \label{eq:GP-var}
    s^2_*(x_*) &\ = C(x_*,x_*)+ \notag  \\ + (\mathbf{h}(x_*)^T &-\mathbf{c}(x_*)^T(\mathbf{C+\Sigma})^{-1}\mathbf{H})^T(\mathbf{H}^T(\mathbf{C+\Sigma})^{-1}\mathbf{H})^{-1}
    (\mathbf{h}(x_*)^T-\mathbf{c}(x_*)^T(\mathbf{C+\Sigma})^{-1}\mathbf{H}).
\end{align}
We note that the mean and kernel functions \emph{interact}: choosing the mean function is analogous to de-trending, and choosing the covariance function is analogous to modeling the residuals. An informative mean function will imply that  the residuals are smaller (lower $\eta^2$) and  de-correlated (small $\theta$'s) compared to assuming a constant mean, which will lead to high $\eta^2$ and larger $\theta$'s.

Within a multi-population model we use a linear mean function to take into account the different trends across populations:
\begin{equation}\label{eq:mean-multi}
    m(x^n)=\beta_0+\beta_1^{ag}x^n_{ag}+\sum_{l=2}^{L}  \beta_{pop,l} x^n_{pop,l}.
\end{equation}
Analogous to the coefficients of categorical covariates in regression, $\beta_{pop,l}$ can be interpreted as the mean difference between log mortality in population $l$ and the baseline. Note that \eqref{eq:mean-multi} implies the \emph{same} shared  Age structure---mortality rates rising exponentially in $x_{ag}$ with slope $\beta_1^{ag}$ in all populations.

\subsubsection*{Observation Likelihood.} We assume a constant observation noise within each population $\sigma_l = \mathrm{StDev}(\epsilon^i) \ \forall i$ in \eqref{eqn:regression} where $x^i_{pop,l} = 1$. This accounts for heterogeneous characteristics when observations from multiple populations are combined, in particular $\sigma_l$ is smaller for larger populations \cite{HLZ20}. The $\sigma_l$'s are estimated via Maximum Likelihood or Markov Chain Monte Carlo along with all other hyperparameters. While assuming homogeneity of noise variance in terms of Age and Year is not entirely realistic, based on the discussion in \cite{LUDKOVSKI2016GAUSSIAN} the impact of more complex observation models is minimal. A common alternative is to assume a Poisson likelihood; however it is well known that mortality data are overdispersed, so a Poisson parametrization is also mis-specified. 

\subsubsection*{Estimating the parameters.}
In single-population GP,  the set of hyper-parameters is $\Theta = (\theta_{ag},\theta_{yr},\eta^2,\sigma^2, \bm{\beta})$. We can learn values of the hyperparameters via optimization of the marginal likelihood function which is the integral of the likelihood times the prior: $p(\mathbf{y|x},\Theta)=\int p(\mathbf{y|f},\Theta)p(\mathbf{f|x},\Theta)d\mathbf{f}$. Since $p(\mathbf{y|x},\Theta)=\mathcal{N}(\mathbf{m,C+\Sigma})$ and if we assume the mean function is known or fixed, the log-likelihood of the marginal  is simply a MVN density:
\begin{equation}\label{eq:log-like}
\log p(\mathbf{y|x},\Theta) = -\frac{1}{2}\mathbf{y}^T(\mathbf{C+\Sigma})^{-1}\mathbf{y}-\frac{1}{2}\log |\mathbf{C+\Sigma}|-\frac{N}{2}\log(2\pi).
\end{equation}
Thus, we have to solve a system of nonlinear equations to maximize \eqref{eq:log-like} which yields the maximum likelihood estimate (MLE). We implement SOGP fitting via the function \code{km()} from the package \code{DiceKriging} \citep{Olivier2012} in \texttt{R}. That package carries out MLE of $\Theta$ using a genetic optimization algorithm. In MOGP with ICM kernel the hyper-parameters are  $\Theta = (\theta_{ag}, \theta_{yr}, (a_{l,q}), (\sigma^2_l), \bm{\beta})$. We use the \texttt{R} package \code{kergp} \citep{Oliver19kergp} to carry out the respective MLE via Kronecker decompositions.

\subsection{Bayesian Gaussian Process Regression}

The GP hyperparameters summarize the covariance structure of the fitted mortality model. The MLE method provides a point estimate $\Theta_{MLE}$ of that structure, i.e.~a ``best guess'' of a GP surface that fits the data. Uncertainty quantification is a major component of our analysis, in particular in assessing how similar or different are the various populations. To this end, we aim to quantify model risk, i.e.~the range of GP models that are consistent with the data via a Bayesian formulation. The Bayesian GP starts with a prior on $\Theta$ and then integrates out the likelihood of the observed data to obtain the posterior distribution of the hyperparameters. A point estimate of $\Theta$ is additionally obtained from the maximum a posteriori (MAP) hyperparameters, $\Theta_{MAP} = \argmax_{\Theta} \sum_i\log p(y_i|\Theta) p(\Theta)$. In fact, MLE can be viewed as a special case of MAP with improper uniform priors. In our analysis,
we employ weakly informative priors to minimize influence of a priori assumptions (so that the data speaks for itself) but still regularize inference by keeping hyperparameters within reasonable ranges.

In practice, computing the posterior density $p(\Theta|\mathcal{D})$ requires to evaluate an intractable multidimensional integral. MCMC algorithms bypass this challenge by drawing samples $\Theta^{(1)},\Theta^{(2)}, \dots,\Theta^{(M)}$ from the posterior. Traditionally, MCMC sampling for GP models was challenging due to strong correlation among the hyperparameters. Recently, powerful new techniques, in particular Hamiltonian Monte Carlo (HMC) have been developed to overcome this challenge. We implement Bayesian GP using \texttt{Stan} \citep{CARPENTER2017} that is built upon efficient HMC. \texttt{Stan} is a free, open-source software, written in \texttt{C++} language, and has risen to be one of the most efficient toolboxes to perform Bayesian inference and optimization for statistical models.

Following \texttt{Stan} recommendations, we standardize the input covariates by subtracting the mean and dividing by the standard deviation, $x^i_{ag,std}:=(x^i_{ag}-\mu_{\mathbf{x}_{ag}})/\sigma_{\mathbf{x}_{ag}}$ to reduce the autocorrelation between the hyperparameters and thus increase the efficiency in the MCMC chains. HMC in \texttt{Stan} further helps to cope with this autocorrelation. \texttt{Stan} returns a set of posterior MCMC samples for $\pmb{\beta}$ and $\Theta$
based on standardized data, so we then have to convert these values back to the original scales. For instance, the sampled hyperparameters $\beta^{std}_\cdot$ of the linear mean function are transformed back by:
\begin{equation*}
\begin{aligned}
    m(x^i) & =\beta_0+\beta_1^{ag}x_{ag}^i
            =\beta_0+\beta_1^{ag}(x_{ag,std}^i\sigma_{\mathbf{x}_{ag}}+\mu_{\mathbf{x}_{ag}})   \\
           & = \big(\beta_0+\beta_1^{ag}\mu_{\mathbf{x}_{ag}}) +
           \beta_1^{ag}\sigma_{\mathbf{x}_{ag}}x_{ag,std}^i  \\
\end{aligned}
\end{equation*}
Thus: $\beta_1^{ag}=\frac{\beta_1^{ag,std}}{\sigma_{\mathbf{x}_{ag}}}$ and $\beta_0=\beta_0^{std}-\bigg(\frac{\beta_1^{ag,std}}{\sigma_{\mathbf{x}_{ag}}}\bigg)\mu_{\mathbf{x}_{ag}}$; in similar fashion, we can transform the lengthscales in the covariance kernel: $\theta_{ag}=\sigma_{\mathbf{x}_{ag}}\theta_{ag}^{std}$ and $\theta_{yr}=\sigma_{\mathbf{x}_{yr}}\theta_{yr}^{std}$. 

\subsubsection*{Bayesian vs MLE MOGP.}

To illustrate uncertainty quantification of a MOGP using a Bayesian \texttt{Stan} framework, we build a joint model on four Male populations from Denmark, France, Sweden and UK. The MOGP model uses the full-rank kernel \eqref{eqn:kernelMulti} trained on Ages 70--84 and Years 1990--2012.
For Bayesian hyper-priors we take $\beta_0 \sim \mathcal{N}(-4,0.5)$, $\beta_1^{ag} \sim \mathcal{N}(0,0.5)$. Inverse-Gamma priors are chosen for the covariance hyperparameters: $\theta^{std}_{ag} \sim \text{Inv-Gamma}(9,12)$, $\theta^{std}_{yr} \sim \text{Inv-Gamma}(9,12)$ which ensures that 99\% of the respective prior is concentrated between 0.01 and 3.3, \citep{Betancourt2017}. For the process variance, we take $\log\eta^2 \sim \mathcal{N}(-3,1)$, and for observation noise $\sigma^2 \sim \mathcal{N}_+(0,0.5)$. Finally, for the population lengthscales in \eqref{eqn:kernelMulti} we use $\log \theta_{l_1,l_2} \sim \mathcal{N}(-1,1)$ for all $l_1, l_2$. (Implementing a \texttt{Stan} model for ICM kernels is beyond the scope of this work.)

Table \ref{tbl:gpB-joint4} in the Appendix reports all the resulting hyperparameters using the \texttt{kergp} engine in \texttt{R} and the \texttt{Stan} HMC. We observe that all MLE fits are within the 95\% posterior credible intervals from the \texttt{Stan} model. Also, the 95\% credible interval for $\beta_1^{ag}$ confirms the significance of the linear effect of Age. Treating Denmark as the baseline country in the mean function, the 95\% CI's of all coefficients $\beta_{pop,l}$'s contain 0, implying that the differences in mortality between Denmark and other populations are not statistically significant. This indicates that there is no clear difference in the respective mortality experience which is intuitive since all  populations are from developed countries within the same geographic area.

\subsection{Performance metrics}
\label{perform-metrics}

To assess model performance we employ two metrics. First, we consider out-of-sample predictive accuracy, comparing realized future mortality to its mean model forecast. The most common choice is root mean squared error (RMSE); however RMSE is highly sensitive to outliers and also to the fact that mortality errors will be necessarily larger at higher Ages due to smaller exposed cohorts. To remedy this, we employ the mean absolute percentage error (MAPE) metric, specifically its symmetric (SMAPE) version that corrects for the tendency of MAPE to put heavier penalties on over-estimating the observations \citep{ARMSTRONG1992, Makridakis1993}:
\begin{equation}
    \text{SMAPE} := \frac{100}{M} \sum_{i=1}^M \frac{|y^i_*-m_*(x^i_*)|}{(|y^i_*|+|m_*(x^i_*)|)/2},
\end{equation}
where $y^i_*$ is the realized observed value at test input $x_*^i$ and $m_*(x^i_*)$ is the predicted log-mortality rate by the model.  Unlike the squared errors, SMAPE is a scale-independent measure that is convenient to compare across different data sets. 

In addition to SMAPE, we also use the Continuous Ranked Probability Score (CRPS) metric to assess the quality of the probabilistic forecasts produced by a MOGP. Indeed, one of the major benefits of GP-based mortality models is a full distribution for future observations $y_*(x_*)$ which allows a more detailed uncertainty quantification beyond just looking at the predictive mean $m_*(x_*)$. CRPS is an example of a proper scoring rule and is defined as
\begin{equation}
    \text{CRPS}(F,y_*) := \int_{\mathbb{R}}\left(F(z)-\mathbbm{1}_{\{z \geq y_* \}} \right)^2 \, dz,
\end{equation}
where $F$ is the predictive (cumulative) distribution and $y_*$ is the realized outcome. Averaging over many outcomes, CRPS can be interpreted as the squared difference between the forecasted and the empirical cumulative distribution functions. In particular, CRPS penalizes both bias and excessive predictive variance. A model with lower mean CRPS is judged to be better.

\textbf{Mortality Improvement Factors.}
A common way to interpret a mortality surface is via the (annual) mortality \emph{improvement factors} which measure longevity changes year-over-year.
In terms of the observations, the raw annual percentage mortality improvement is
    $1-\frac{\exp\big(y(x_{ag};x_{yr})\big)}{\exp\big(y(x_{ag};x_{yr}-1)\big)}$.
The smoothed improvement factor is obtained by replacing $y$'s by the GP model posterior $m_{*}$'s: 
\begin{equation}\label{eq:mi}
    \partial m^{GP}_{back}(x):=\Bigg[1-\frac{\exp(m_{*}(x_{ag};x_{yr}))}{\exp(m_{*}(x_{ag};x_{yr}-1))}\Bigg].
\end{equation}

\section{Maximizing Predictive Accuracy through Joint Modeling}\label{sec:smape}

We begin our illustrations by showcasing the improved predictive accuracy available from fusing data from 2 populations. Our first case study includes Male mortality modeling across Sweden and Denmark ($l=1$: Denmark and $l=2$: Sweden). The two countries share similar demographic characteristics,  and are Nordic neighbors. 
Our mean function takes into account the separation in mortality between Denmark and Sweden:
\begin{equation}\label{eq:mean-multi-2}
    m(x^n)=\beta_0+\beta_1^{ag}x^n_{ag}+ \beta_{pop,2} x^n_{pop,2}.
\end{equation}
Analogous to a coefficient of categorical covariates in regression, $\beta_{pop,2} \equiv \beta_{SWE}$ can be interpreted as the mean difference between log mortality in Sweden and in the baseline country, Denmark. 
Our second case study looks at joint Male/Female modeling for Denmark; in that case $\beta_{pop,2} \equiv \beta_{FEM}$ is the mean difference between female log mortality and male log mortality.

\begin{table*}[t]
\centering
\renewcommand{\arraystretch}{1.25}
\captionsetup{justification=centering}
\caption{Prediction accuracy via SMAPE for SOGP and 2-population Full rank models. The test set is Ages 70--84 in Years 2013, 2015, and 2016.}
\label{tbl:smape-mort}
\resizebox{\columnwidth}{!}{\begin{tabular}{cccccccccc}\toprule
\multicolumn{2}{c}{\multirow{2}{*}{SMAPE}} & \multicolumn{2}{c}{2013 (one-year out)} & & \multicolumn{2}{c}{2015 (three-year out)} & & \multicolumn{2}{c}{2016 (four-year out)} \\ \cline{3-4} \cline{6-7} \cline{9-10}
\multicolumn{2}{c}{}  & SOGP & MOGP & & SOGP & MOGP  & & SOGP & MOGP    \\ \hline
\multirow{2}{*}{Age $\in [70,84]$ } & Denmark & 1.5798 & \textbf{1.4451} & &  1.3445 & \textbf{1.2862} &  &  1.2584    &  \textbf{1.1955}    \\
& Sweden & 1.0450 & \textbf{0.8256}  & &  1.9752 & \textbf{1.1011} & & 2.5272   &  \textbf{0.9038}  \\ \hline
\multicolumn{2}{c}{}  & SOGP & MOGP & & SOGP & MOGP  & & SOGP & MOGP    \\ \hline
\multirow{2}{*}{ Age $\in [70,84]$} & Female &  0.9422 &    \textbf{0.8834} & &  1.8973 & \textbf{1.7845} & & 1.4010 & \textbf{1.2269}  \\
& Male &  1.5802 & \textbf{1.5062} & & 1.3444 & \textbf{1.2454} & &  1.2583 & \textbf{1.1819}    \\  \bottomrule
\end{tabular}}
\end{table*}

\begin{figure}[!ht]
    \centering
    Denmark \\
    {{\includegraphics[width=0.31\textwidth]{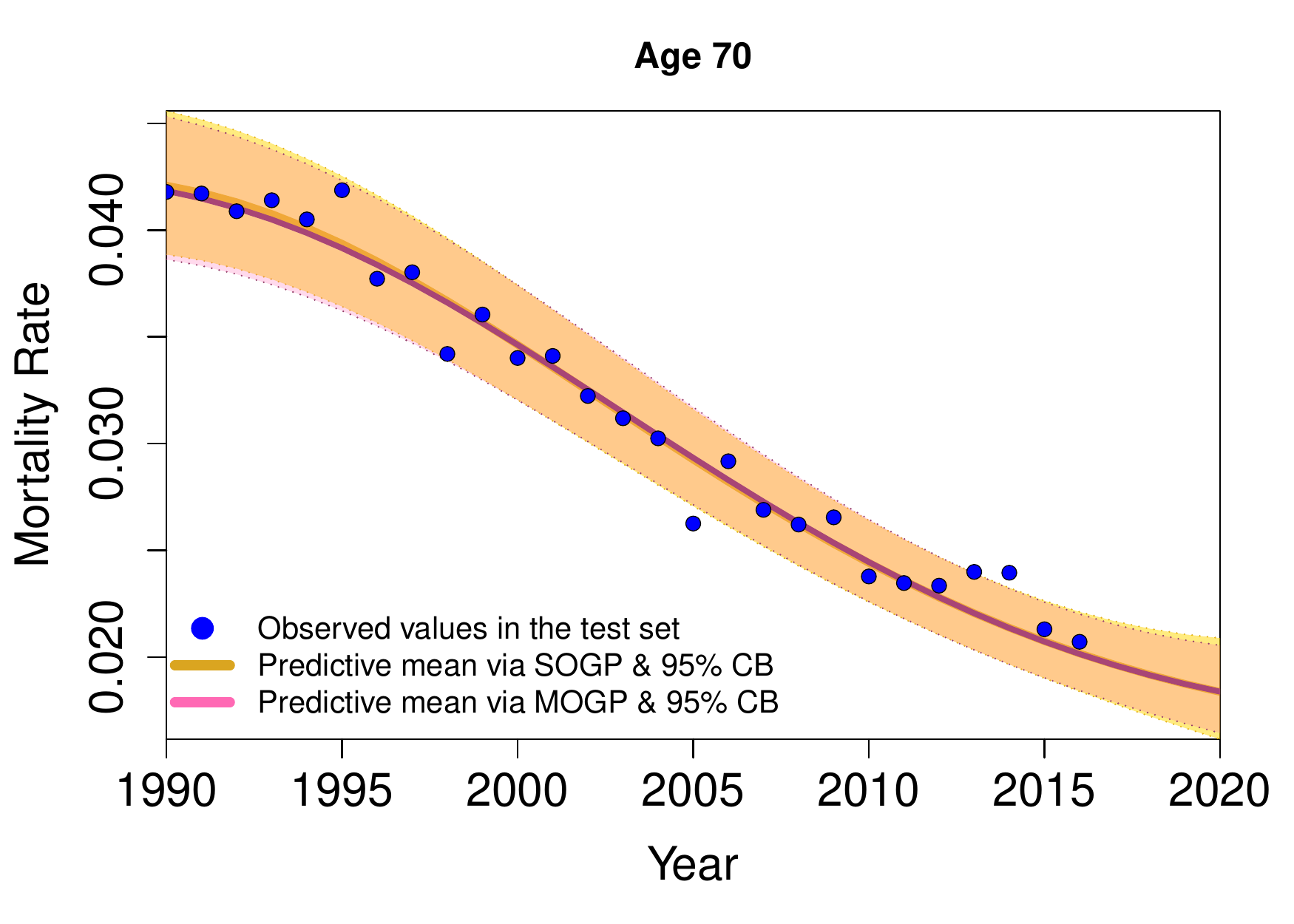}}}
    {{\includegraphics[width=0.31\textwidth]{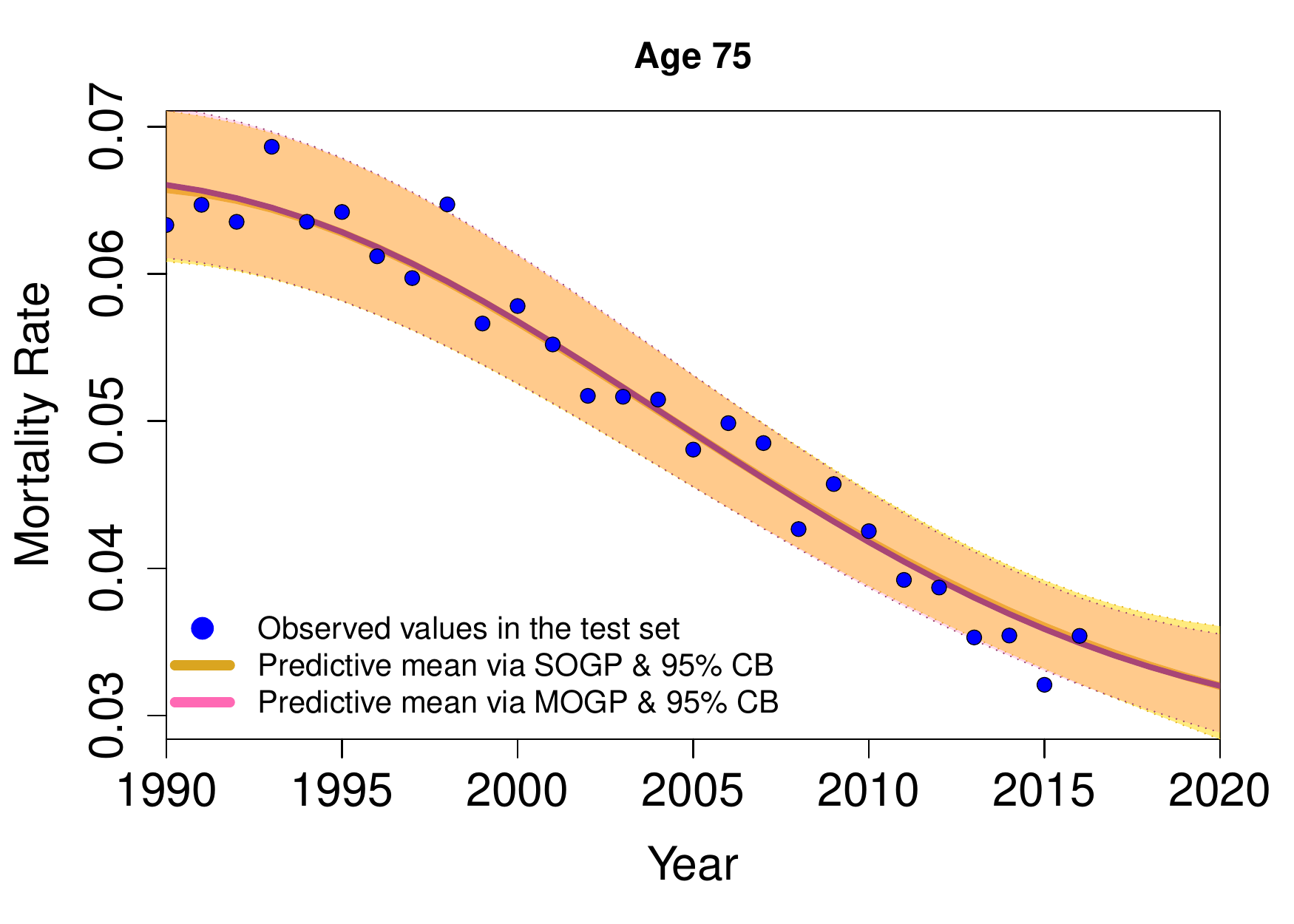}}}
    {{\includegraphics[width=0.31\textwidth]{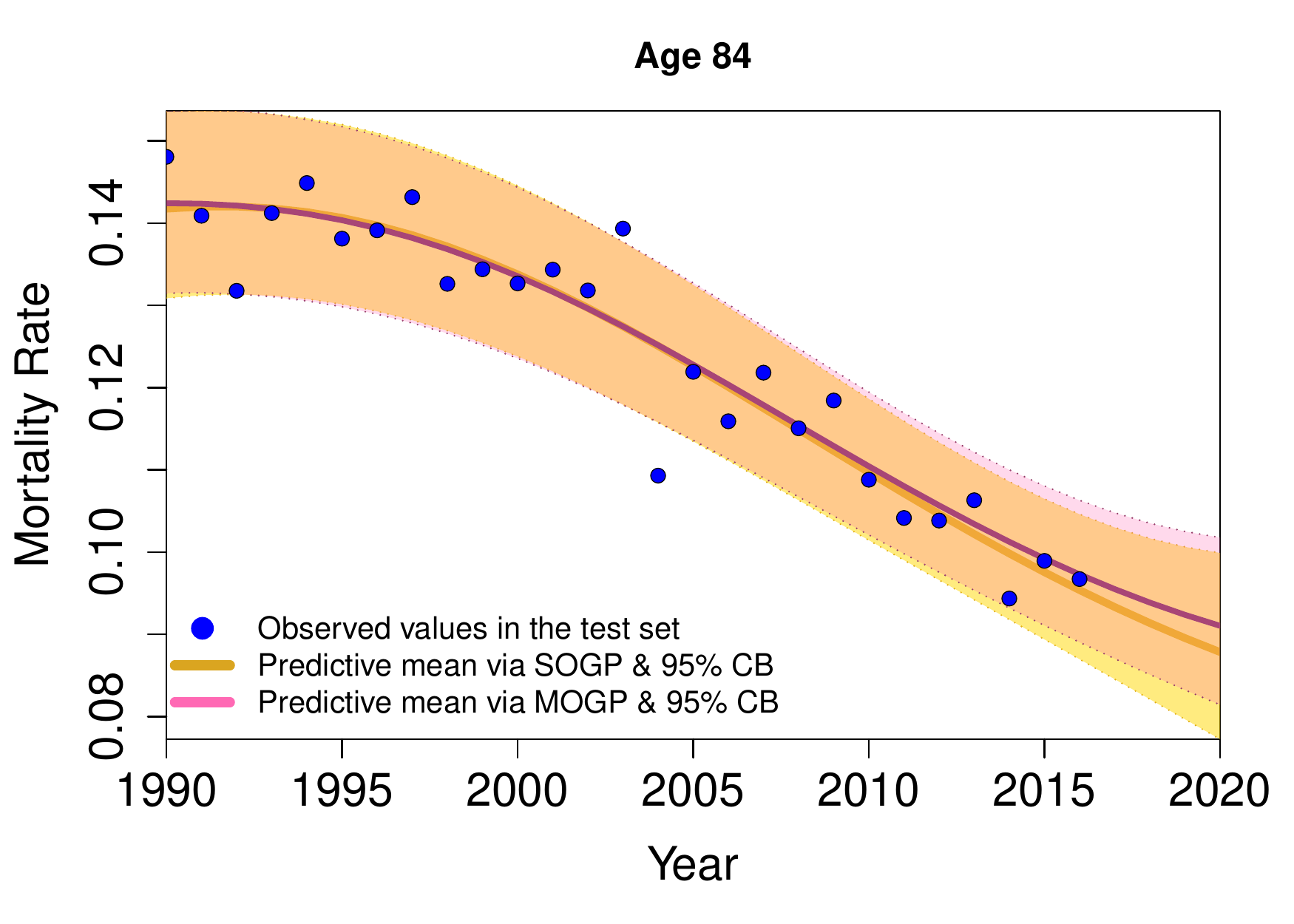}}}\\
    Sweden \\
    {{\includegraphics[width=0.31\textwidth]{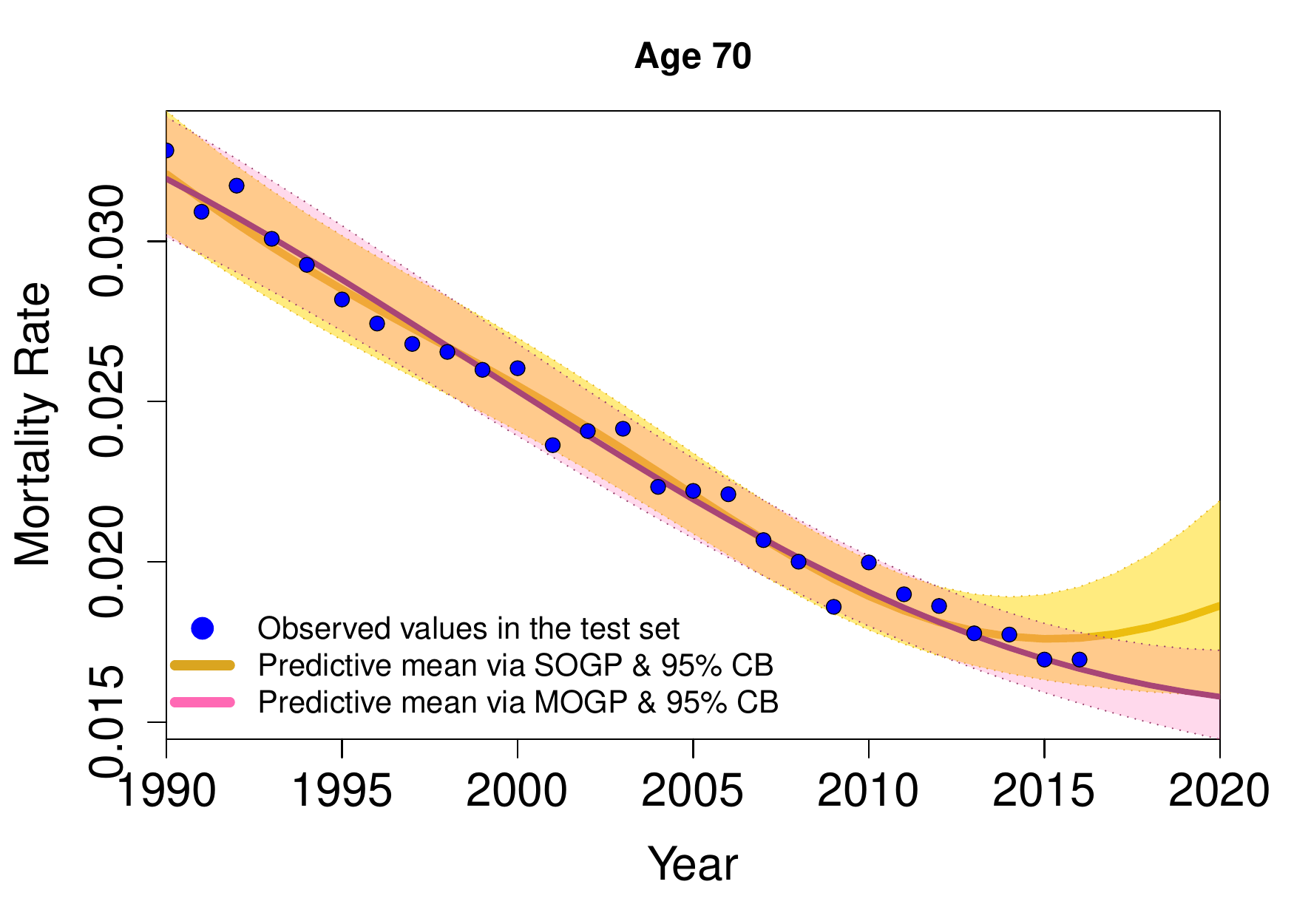}}}
    {{\includegraphics[width=0.31\textwidth]{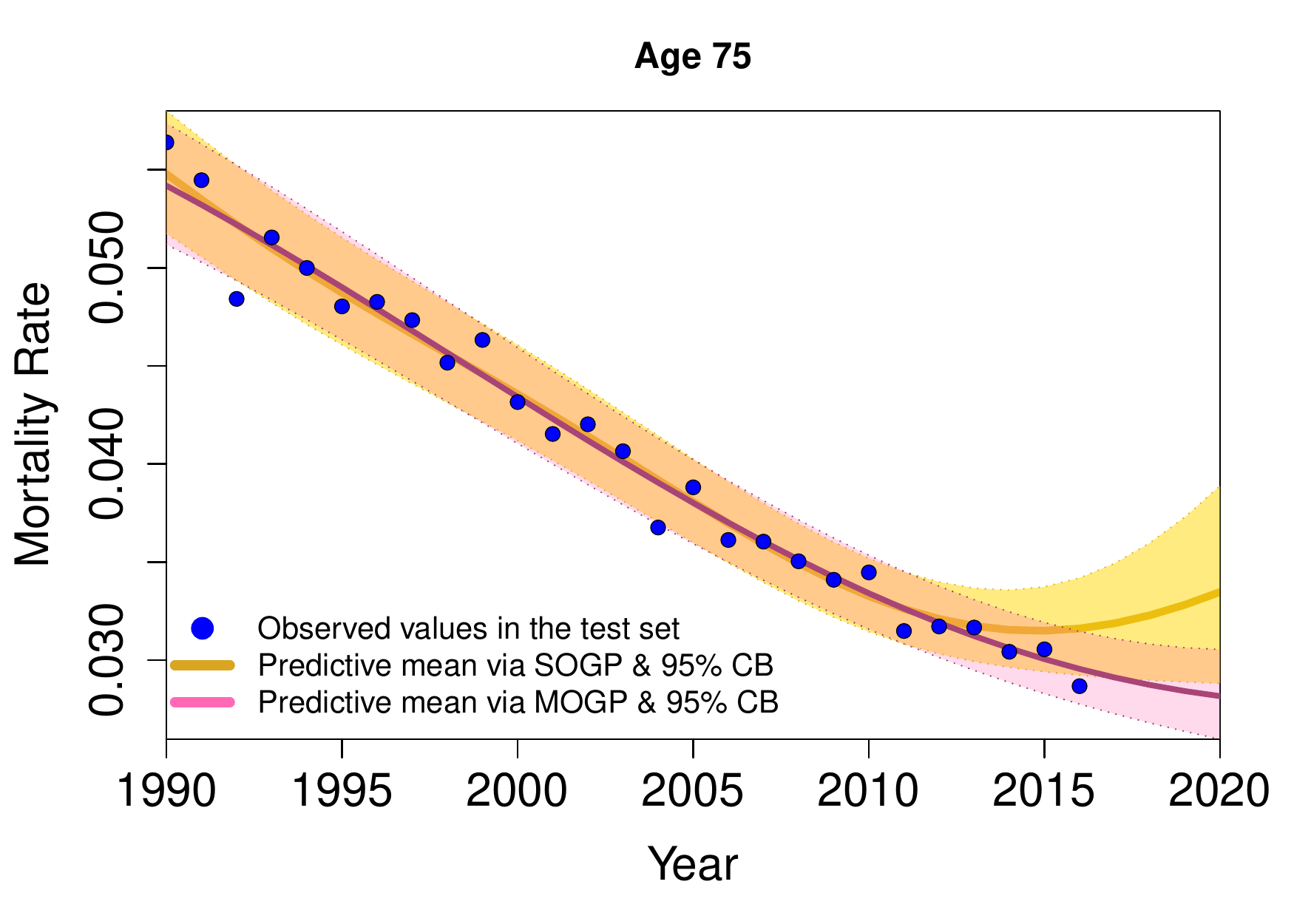}}}
    {{\includegraphics[width=0.31\textwidth]{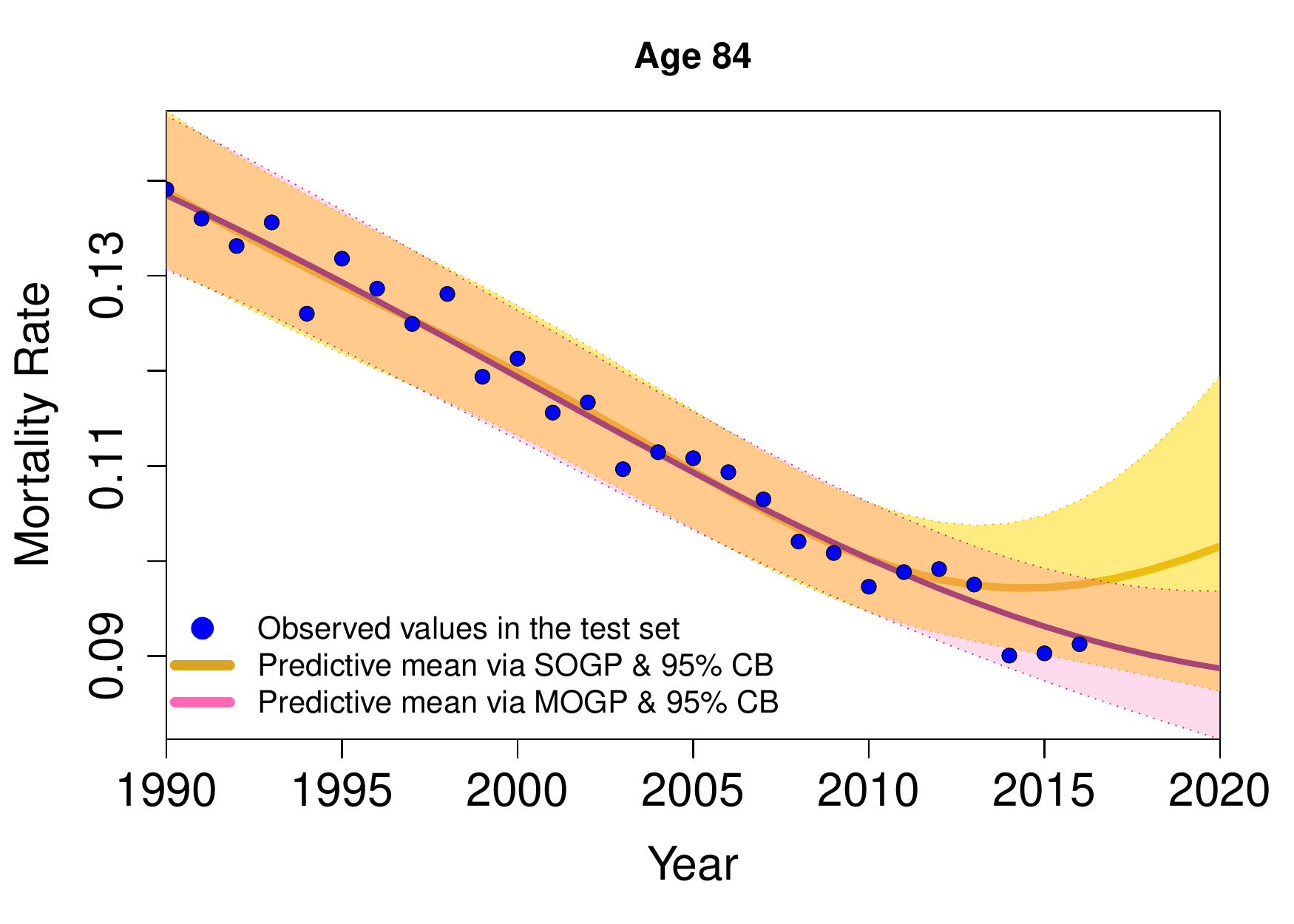}}} \\
    \caption{95\% credible intervals for observed log-mortality $y(x_*)$ across the individual SOGP and Full MOGP models. Top row: Denmark Males; bottom row: Sweden Males. Up to 2011, the smoothed mortality curves and CIs are essentially identical.  \label{fig:95cbands-ys}}
\end{figure}

Figure~\ref{fig:95cbands-ys} shows the raw observations together with the GP-based predictive intervals for the first case study. Specifically, we plot the 95\% predictive credible bands for $y_*(x_*)$ for three representative ages. The forecast period includes both in-sample (1990-2012) and out-of-sample (2013-2020). We observe that while for Denmark the diffferences between SOGP and MOGP forecasts  are very slight, in Sweden the two forecasts differ noticeably out-of-sample. Table~\ref{tbl:smape-mort} compares the predictive accuracy between the models and indicates that the MOGP forecast is more accurate (smaller SMAPE) in both populations.

\begin{figure}[!t]
    \centering
    \captionsetup{justification=centering}
    \subfloat[Denmark Males]{{\includegraphics[width=7.5cm]{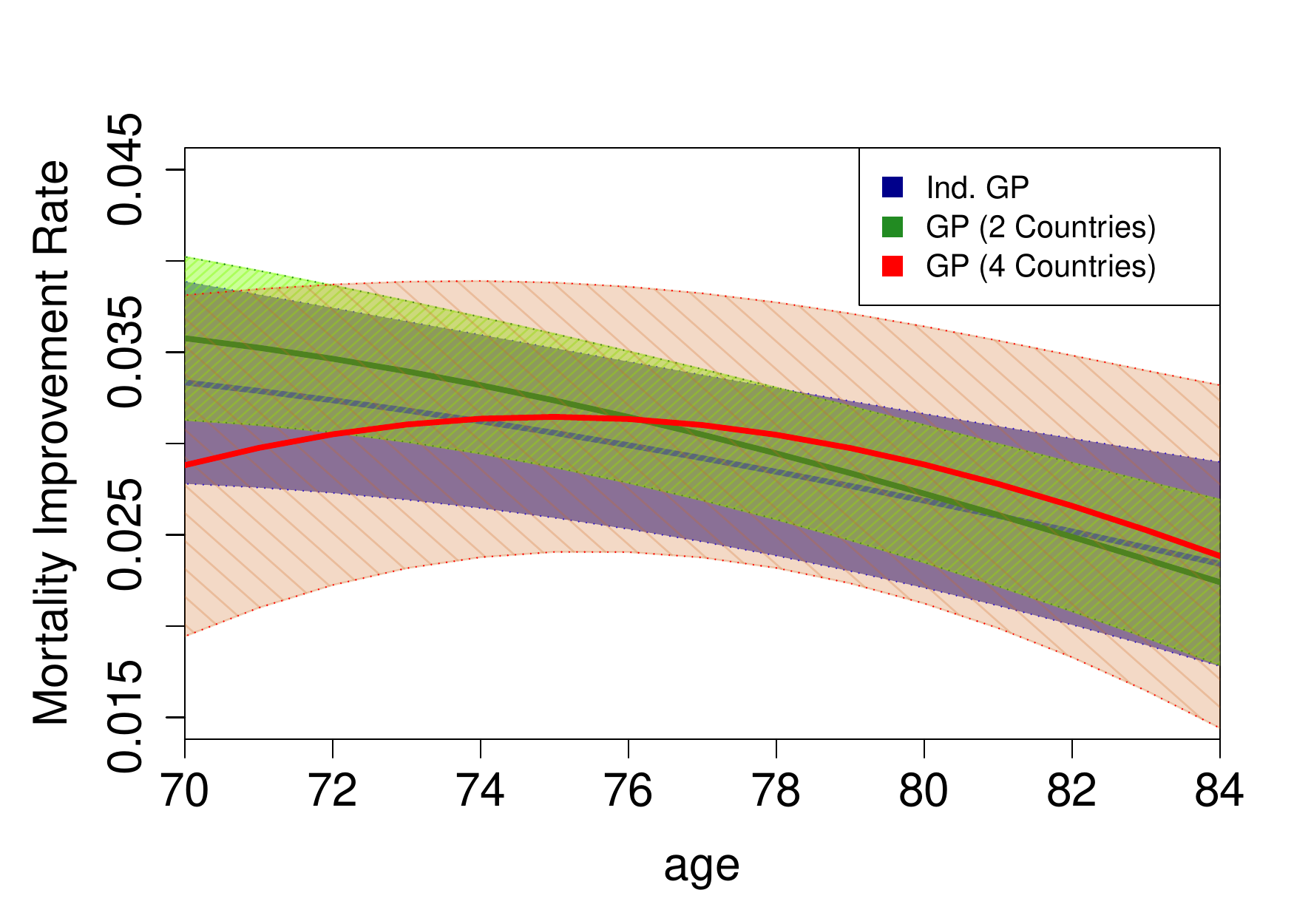}}}
    \qquad
    \subfloat[Sweden Males]{{\includegraphics[width=7.5cm]{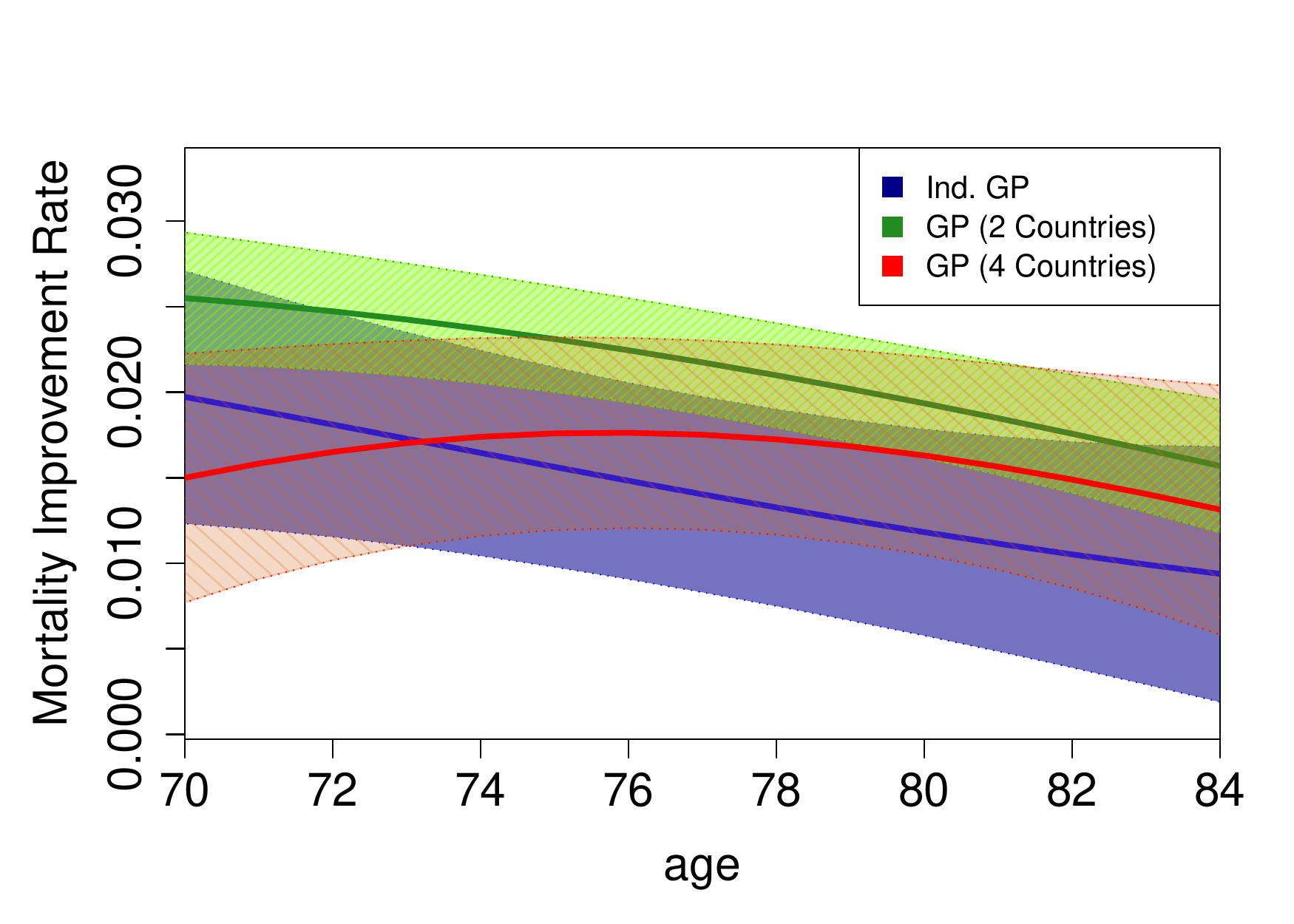}}}
    \qquad
    \caption{Comparison of annual mortality improvement factors between different joint models. Besides the mean improvement factors $\partial m^{GP}_{back}(ag;2012)$ \eqref{eq:mi}  for  Ages $70,\ldots,84$, we also show the respective 95\% posterior credible band. }
    \label{fig:cbands95-mi}
\end{figure}

To highlight further differences between SOGP and Full MOGP,  Figure \ref{fig:cbands95-mi} examines the respective predicted annual mortality improvement factors $\partial m^{GP}_{back}(x)$. We compare SWE and DEN SOGPs against a 2-population DEN+SWE MOGP, and a 4-population DEN+FRA+GBR+SWE MOGP model, cf.~Table~\ref{tbl:gpB-joint4} in the Appendix.
Large $\theta_{ag}$ lengthscales in SOGP models lead to essentially linear improvement rate factors (blue curves). In the SWE + DEN MOGP  (green curves), the Age lengthscale decreases and $s_*(x_*)$ falls, so the improvement rate factors become more Age-dependent and with tighter credible bands. This effect becomes even stronger with four populations. The corresponding smoothed curves (colored in red) are quite nonlinear, and in particular imply that improvement at young Ages ($<60$) has slowed dramatically. This illustrates that a joint model is better able to distinguish between signal and ``noise'' and therefore pick up divergent changes in mortality faster, while a single-population model would often smooth latest changes away.

\subsection{Selecting Populations for a Joint Model}
\label{cluster}

Intuitively, incorporating more information from different populations through a MOGP ought to produce more accurate predictions and reduce predictive uncertainty. To visualize how increasing $L$ affects the changes in SMAPE and CRPS, Table~\ref{tbl:manyPops-quality} reports the 3-year average improvement in these metrics as $L$ varies from 2 to 6. (ICM ranks $Q=Q(L)$ were selected each time using BIC). Overall, we observe that information fusion is very helpful for Hungary, but not as much for UK. This links to the respective credibility of the target populations: observation noise $\sigma_l^2$ is large in Hungary but low in UK, so additional data will benefit the former more.

\begin{table}[ht]
\centering
\caption{Prediction quality via 3-year average improvement in SMAPE and CRPS in  ICM models with $L=2,\ldots,6$ populations. The baseline models are SOGP for UK and Hungary, respectively.
\label{tbl:manyPops-quality}}
\begin{tabular}{lrrlrr} \toprule
 & \begin{tabular}[c]{@{}r@{}}Improvement\\  in SMAPE (\%)\end{tabular} & \begin{tabular}[c]{@{}r@{}}Improvement \\ in CRPS (\%)\end{tabular} &  & \begin{tabular}[c]{@{}r@{}}Improvement \\ in SMAPE (\%)\end{tabular} & \begin{tabular}[c]{@{}r@{}}Improvement \\ in CRPS (\%)\end{tabular}  \\ \midrule
UK + 1 & $-$2.805  & 1.834 & Hungary + 1 & 7.501 & 8.293 \\
UK + 2 & 1.189 & 3.218 & Hungary + 2 & 1.839 & 8.543    \\
UK + 3 & 3.995 & 3.279 & Hungary + 3 & 3.913 & 7.218    \\
UK + 4 & 1.778 & 0.933 & Hungary + 4 & 2.677 & 15.559  \\
UK + 5 & 1.764 & $-$4.271 & Hungary + 5 & 10.585 & 13.345 \\
\bottomrule
\end{tabular}
\end{table}

From a complementary perspective, Table~\ref{tbl:corr-study} shows how the inferred cross-population correlations change as we add a new populations. We report results both for a full-rank MOGP and for ICM with $Q=2,3,4$. We observe that the correlation matrix is generally stable, although some correlations
can be quite different moving from one rank to another. In the AUT-SUI-GBR model, the correlation between Switzerland and UK is $r_{SUI,GBR}=0.76$ when $Q=2$, but rises to 0.88 when $Q=3$. We further note a broad agreement between the correlation structure learned with an ICM and Full-rank MOGP kernels.

\begin{table}[ht]
\centering
\caption{Cross-correlation among 2-4 populations in MOGP models. All models are fitted on Ages 70--84 and reported values are averages over 3 training sets covering 1990 through 2014--2016. Italics indicate the ICM model with the smallest BIC.
\label{tbl:corr-study}}
\begin{tabular}{rrrrr} \toprule
\multicolumn{1}{l}{}         & Full rank & ICM ($Q=2$) & ICM ($Q=3$)     & ICM ($Q=4$) \\ \midrule
\multicolumn{1}{l}{Austria, UK}                       &           &             &                 &             \\
$r_{AUT, ~GBR}$                 & 0.8432    & 0.8612      &                 &             \\ \midrule
\multicolumn{2}{l}{Switzerland, UK}                              &             &                 &             \\
$r_{SUI, ~GBR}$                  & 0.8535    & 0.8645      &                 &             \\ \midrule
\multicolumn{2}{l}{Austria,  Switzerland, UK}                    &             &                 &             \\
$r_{AUT, ~SUI}$                    & 0.9151    & 0.9677      & \textit{0.9680} &             \\
$r_{AUT, ~GBR}$                     & 0.8590    & 0.8968      & \textit{0.8460} &             \\
$r_{SUI, ~GBR}$                     & 0.8514    & 0.7585      & \textit{0.8841} &             \\ \midrule
\multicolumn{2}{l}{Austria, Germany, Switzerland, UK}            &             &                 &             \\
$r_{AUT, ~GER}$                      & 0.9570           & 0.9991      & \textit{0.9869} & 0.9888      \\
$r_{AUT, ~SUI}$                      & 0.9280          & 0.9956      & \textit{0.9639} & 0.9658      \\
$r_{AUT, ~GBR}$                       & 0.8730          & 0.8330      & \textit{0.8917} & 0.8827      \\
$r_{GER, ~SUI}$                       & 0.9047          & 0.9986      & \textit{0.9395} & 0.9447      \\
$r_{GER, ~GBR}$                        & 0.8504          & 0.8091      & \textit{0.8340} & 0.8267      \\
$r_{SUI, ~GBR}$                        &0.8674           & 0.7792      & \textit{0.9047} & 0.9187     \\ \midrule
\end{tabular}
\end{table}

\begin{figure}[!ht]
    \centering
    \begin{tabular}{lr}
    \begin{minipage}{0.4\textwidth}
    \begin{tabular}{lcc} \toprule
        & $a_{l,1}$     & $a_{l,2}$     \\ \midrule
        $l = 1$: EST  & \textbf{0.1925} & 0.0815 \\
        $l = 2$: HUN  & \textbf{0.1692} & 0.0673 \\
        $l = 3$: LTU  & \textbf{0.1757} & 0.0411 \\
        $l = 4$: NED  & 0.0846 & \textbf{0.1127} \\
        $l = 5$: POL  & \textbf{0.1670} & 0.0886 \\
        $l = 6$: SWE  & 0.0614 & \textbf{0.0943} \\
        $l = 7$: SUI  & 0.0745 & \textbf{0.1113} \\
        $l = 8$: GBR   & 0.1118 & \textbf{0.1253} \\ \bottomrule
        \end{tabular} \end{minipage} &
    \begin{minipage}{0.45\textwidth}
    \includegraphics[scale=0.45]{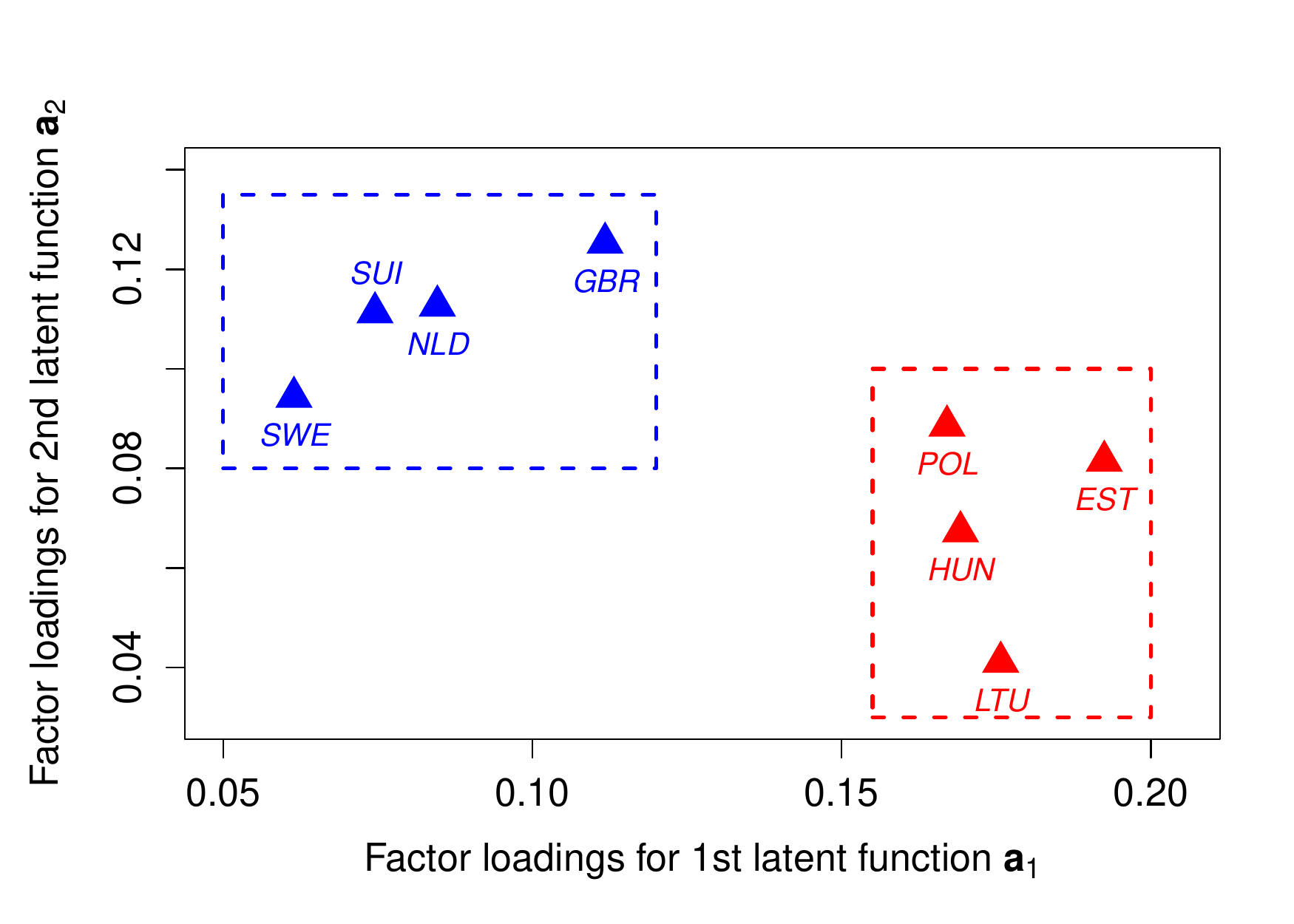}\end{minipage} \end{tabular}
    \captionlistentry[table]{}
    \captionsetup{labelformat=andtable}
    \caption{Factor loadings $a_{l,q}$ in the 8-population ICM with $Q=2$ in Table~\ref{tbl:fullvsICM} - Case study II.\label{fig:loadings}}
\end{figure}

The factor loadings ($a_{l,q}$'s) in ICM provide insight regarding the dependence patterns across populations. The interpretation of ICM loadings is analogous to Principal Component Analysis when attempting to describe the data through independent transformed latent functions. For example, we consider factor loadings in case study II in Table \ref{tbl:fullvsICM}. The best ICM kernel has rank $Q=2$ suggesting that two latent factors are sufficient to explain variation over the eight countries considered. In fact, the first latent component is strongly correlated with Eastern-Central European countries  while the second factor is the major contributor to the Western European population. This interpretation helps us identify two well-separated clusters among these eight countries, visualized in Figure~\ref{fig:loadings} by plotting the ICM loadings $a_{l,2}$ against $a_{l,1}$. Note that these factor loadings can be translated into correlation and imply that Hungary is more correlated with members in the same cluster and less correlated with Western European populations.

\subsection{Incorporating Latest Data from Other Populations}
\label{latestData}

In HMD, the reported data from different countries arrives non-synchronously. Indeed, the last available year of data varies from one country to another. The prevailing approach is to consider the time period that is common to all countries that are being modeled. This implies that the most recent observations may be dropped for some countries. Of course, such recent data is in fact the most informative for picking up new insights about the present longevity trend. Note that the HMD datasets are updated continuously, so that which datasets have the latest observations changes dynamically over time.

To assess the value of information fusion and its link to population cross-correlation, we investigate the improvement in prediction in MOGP over SOGP.
 To do so, we set up a ``notched" 2-population training set where the foreign population has one more year of data and the assessment is based on one-year-ahead prediction for the domestic target population. Note that such ``notch'' extrapolation is not possible in the Lee-Carter framework that requires rectangular datasets.
All ICM MOGP models take $L=2$ and are fitted on Ages 70--84 in three different time frames: period 1990--2013/2014 for 2014 forecast, 1990--2014/2015 for 2015 forecast, and 1990--2015/2016 for 2016 forecast. The comparator training datasets for SOGP models did not  have mortality information in the calendar year of forecast (e.g.~training on 1990--2013 for 2014 forecast). We report the resulting 3-year average percent improvement in CRPS and SMAPE between MOGP and SOGP. The positive average improvement is equivalent to MOGP models having smaller SMAPE and CRPS.

\begin{figure}[!t]
\centering
    \subfloat[Hungary: $\sigma^2_l =\num{2.083e-03}$]{{\includegraphics[width=5.25cm]{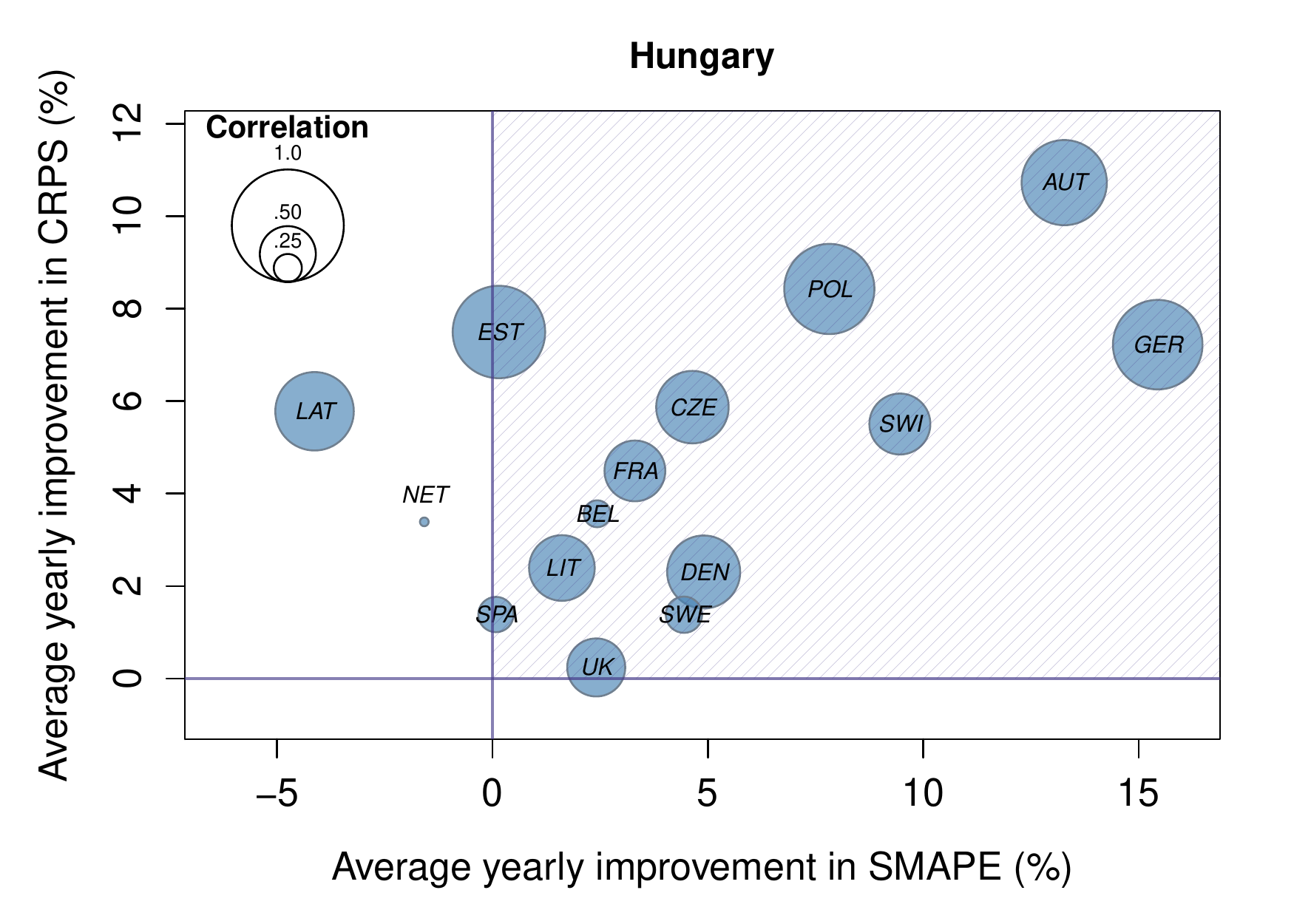}}}
     \subfloat[UK: $\sigma^2_l=\num{6.834e-04}$]{{\includegraphics[width=5.25cm]{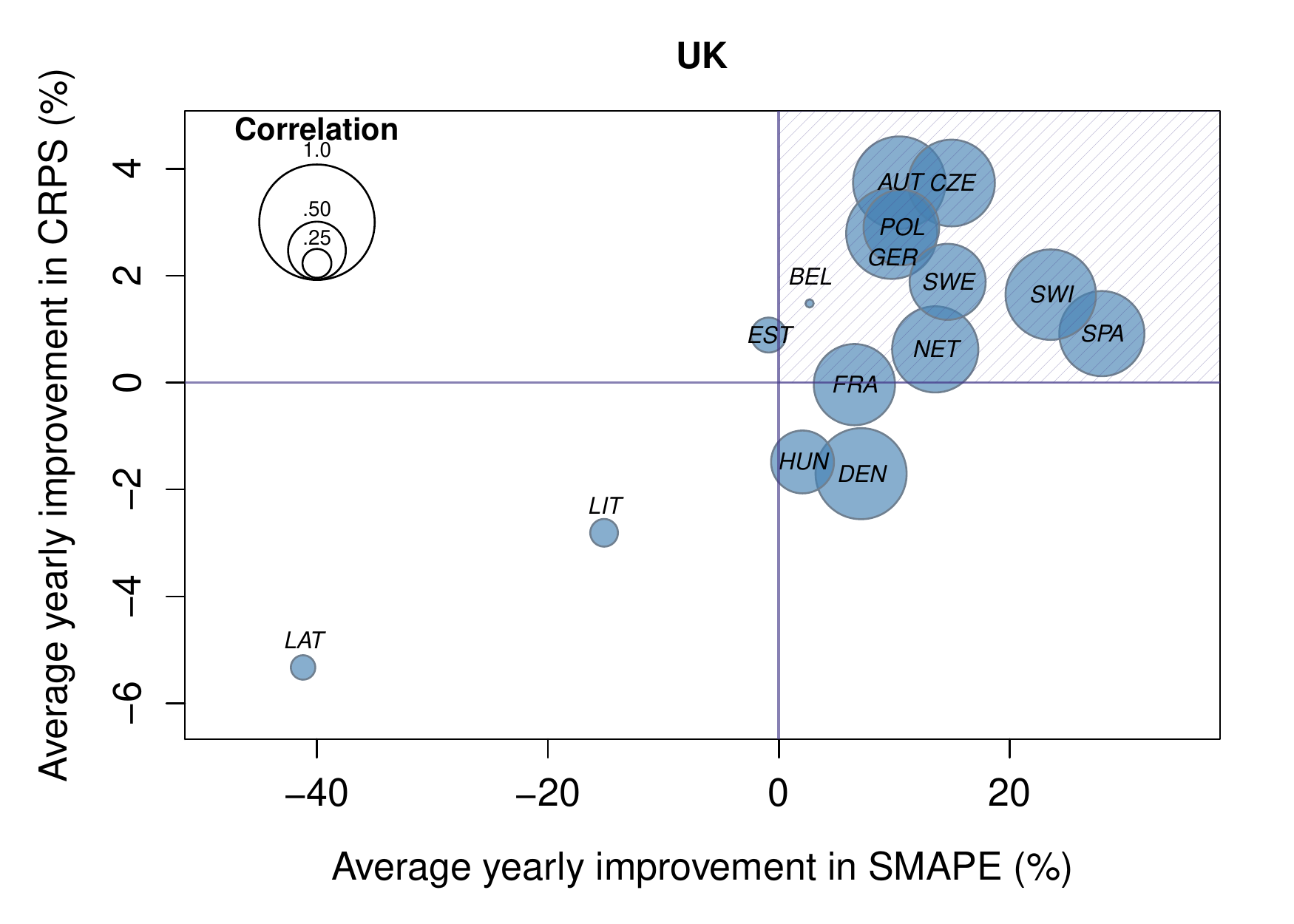}}}
    \subfloat[Linear slope of Improvement in SMAPE vs. correlation]{{\includegraphics[width=5.25cm]{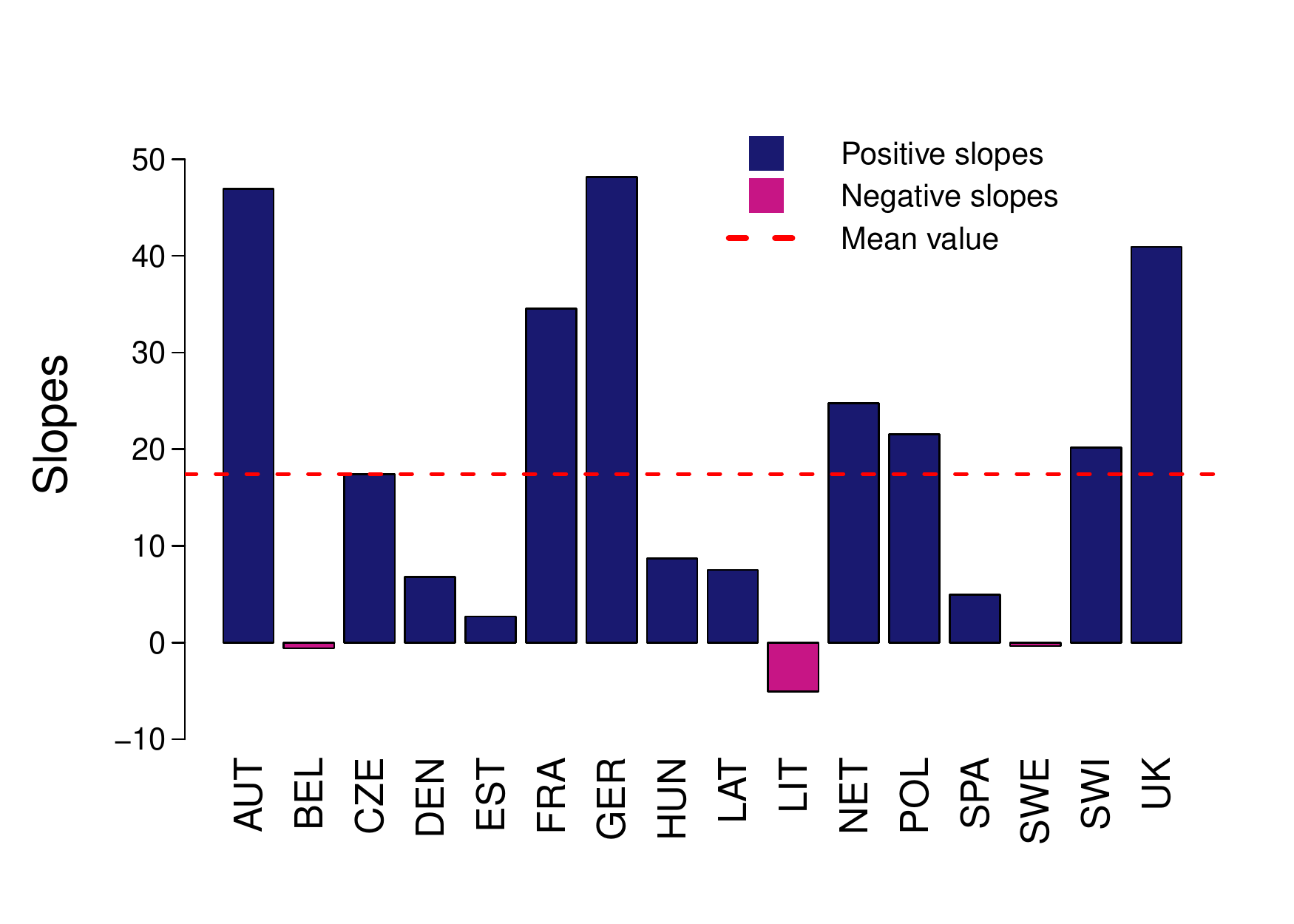}}}
    \caption{Prediction improvements for Hungarian \& UK Males and the impact of cross-correlation on the improvement in SMAPE across 16 European national populations.}
    \label{fig:mogp_impr_noise}
\end{figure}

Figure \ref{fig:mogp_impr_noise} displays the improvement in SMAPE and CRPS of two-population MOGPs vs SOGP. We plot the results against the correlation $r_{l_1, l_2}$ between the two populations modeled in MOGP and consider two randomly chosen target populations: Males in Hungary and Males in UK. Shaded regions indicate which MOGP models have both SMAPE and CRPS values less than the baseline. We observe that joint modeling generally yields higher improvement in Hungary compared to UK which is driven by the former's relatively smaller population which translates into larger $\sigma_l$ and more opportunity for information fusion. Thus, for UK the single-population model is often competitive in its forecasting performance with a two-population MOGP.

The above discussion suggests that fusing highly correlated populations is better for predictive accuracy. To confirm this hypothesis,  Figure \ref{fig:mogp_impr_noise}.(c)
summarizes the relationship between SMAPE improvement and correlation $r_{l_1,l_2}$. We used all 16 populations as targets and built $16 \times 15$ two-population MOGPs to record the resulting MOGP-SOGP improvements in SMAPE like in the right and middle panels. For each target population, we then fitted a linear regression model treating the 3-year average improvement in SMAPE as the dependent outcome and the correlation as the independent variable: SMAPE CHANGE $= b_0 + b_1 r_{l_1, l_2}$. Figure \ref{fig:mogp_impr_noise}.(c) displays the resulting slopes $b_1$ across the 16 populations. Positive $b_1$ implies that higher correlation leads to lower SMAPE, i.e.~higher predictive accuracy. The mean value of the $b_1$-slopes is highly positive and is around 20\%. These empirical features suggest that one should indeed focus on aggregating \emph{related} populations and discard unrelated ones. This is consistent with the results in Table \ref{tbl:fullvsICM} earlier. Most populations in Case Study II are less correlated with Hungary compared to Case Study I and as a result the SMAPE improvement in Case Study I is higher (10.75\%) compared to Case Study I (-2.77\%).

\begin{figure}[!ht]
    \centering
    \begin{tabular}{ccc}
    {{\includegraphics[width=0.31\textwidth,trim=0in 0.1in 0in 0.1in]{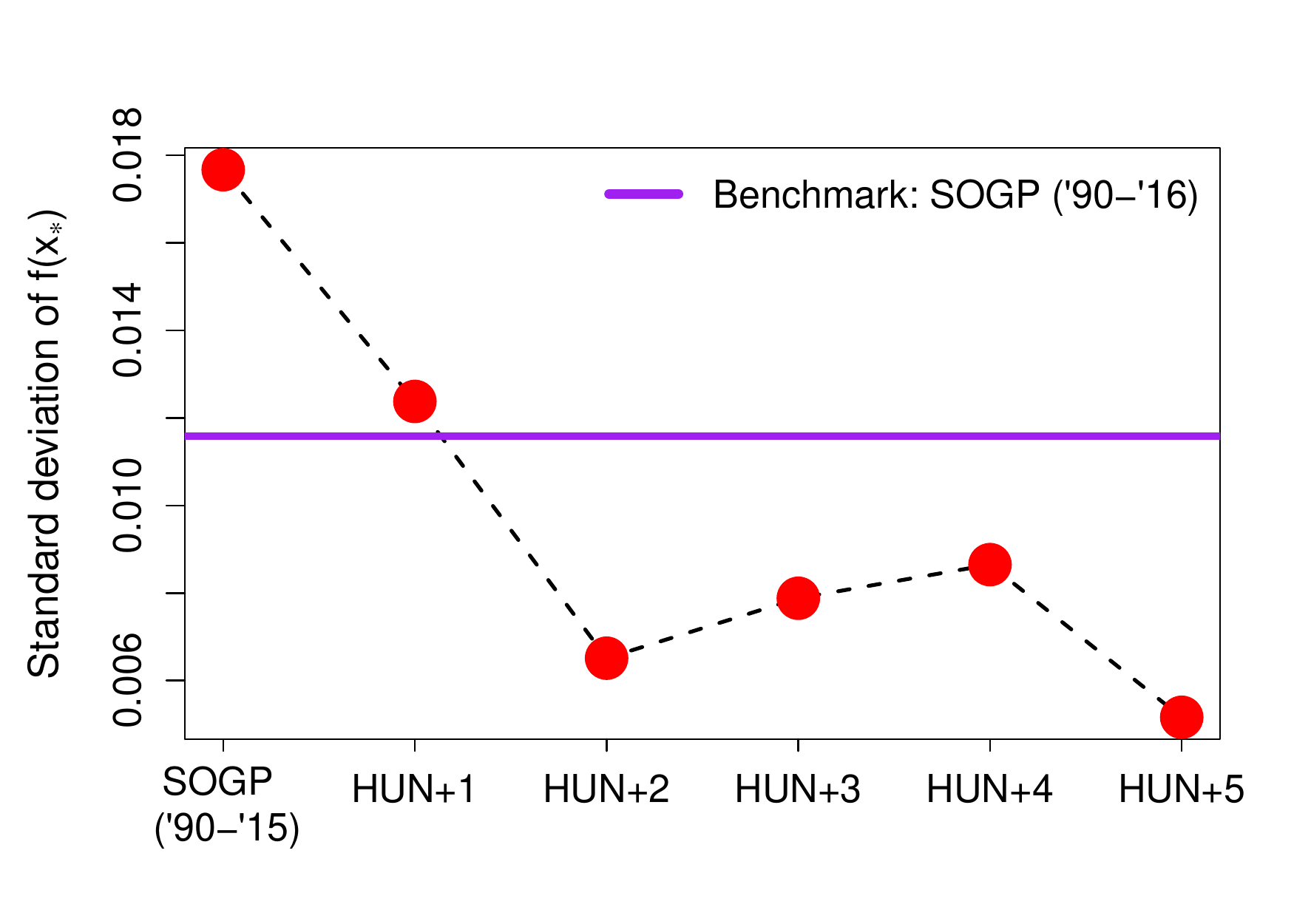}}} &
    {{\includegraphics[width=0.31\textwidth,trim=0in 0.1in 0in 0.1in]{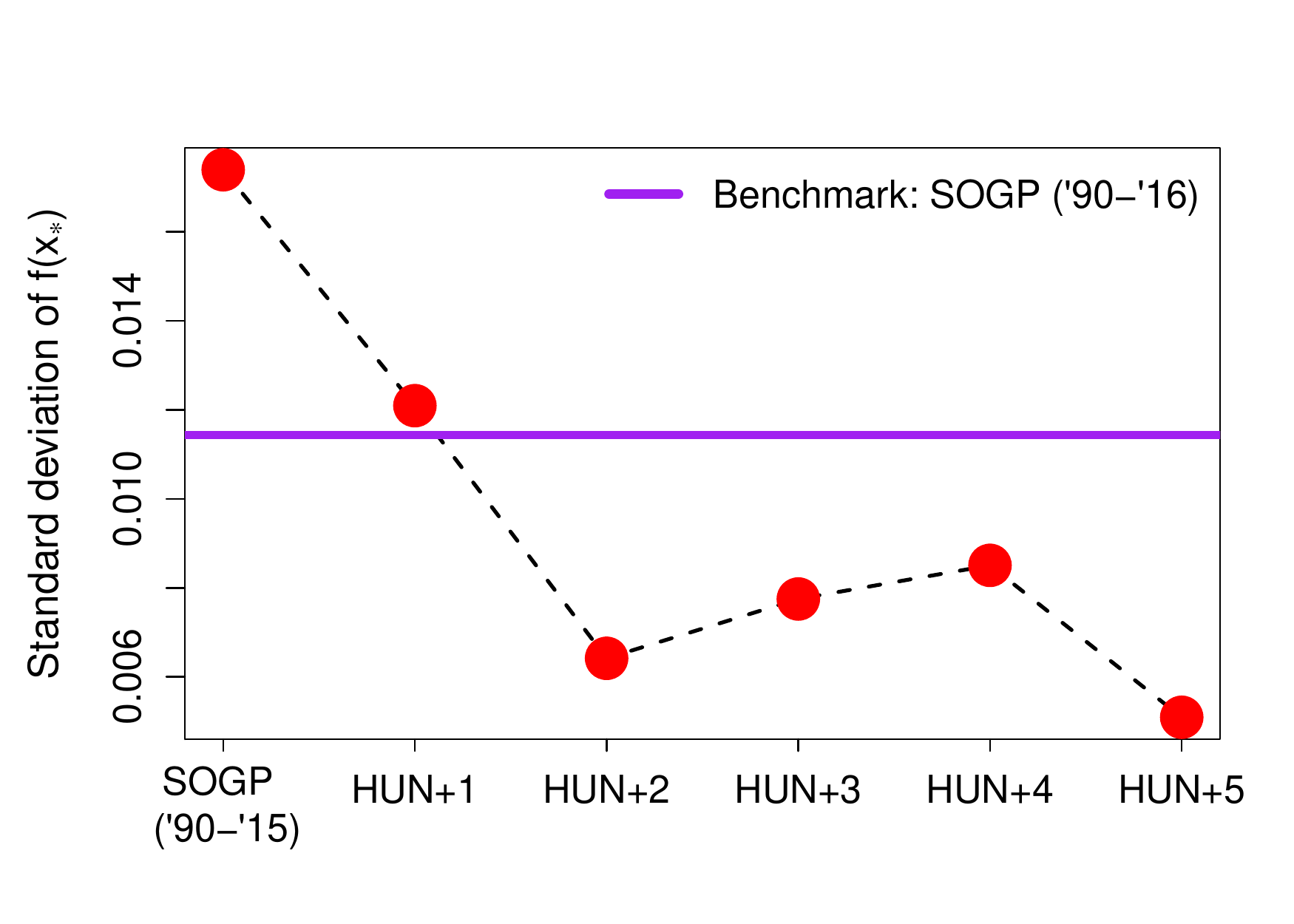}}} &
    {{\includegraphics[width=0.31\textwidth,trim=0in 0.1in 0in 0.1in]{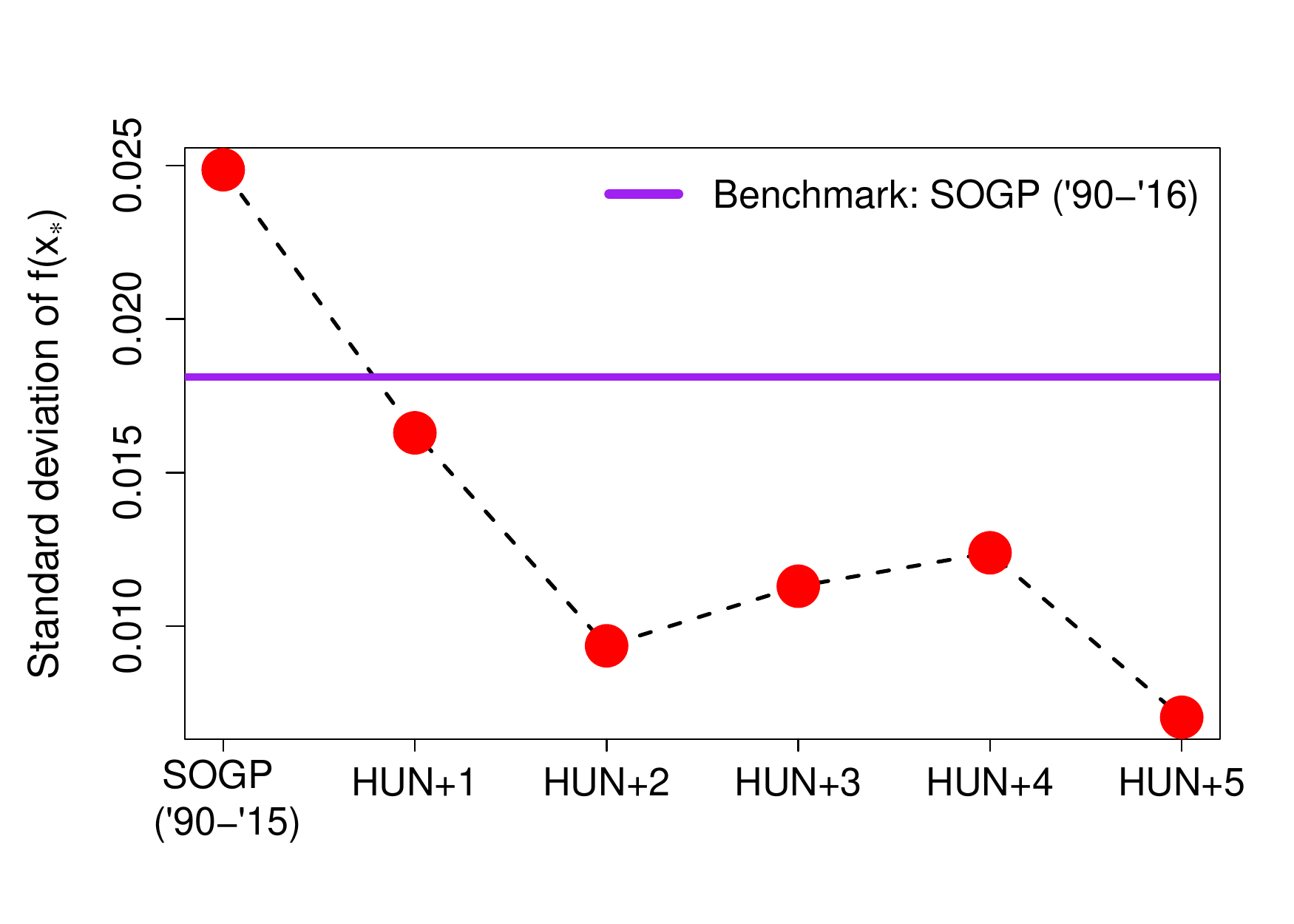}}}\\
    {{\includegraphics[width=0.31\textwidth,trim=0in 0.1in 0in 0.1in]{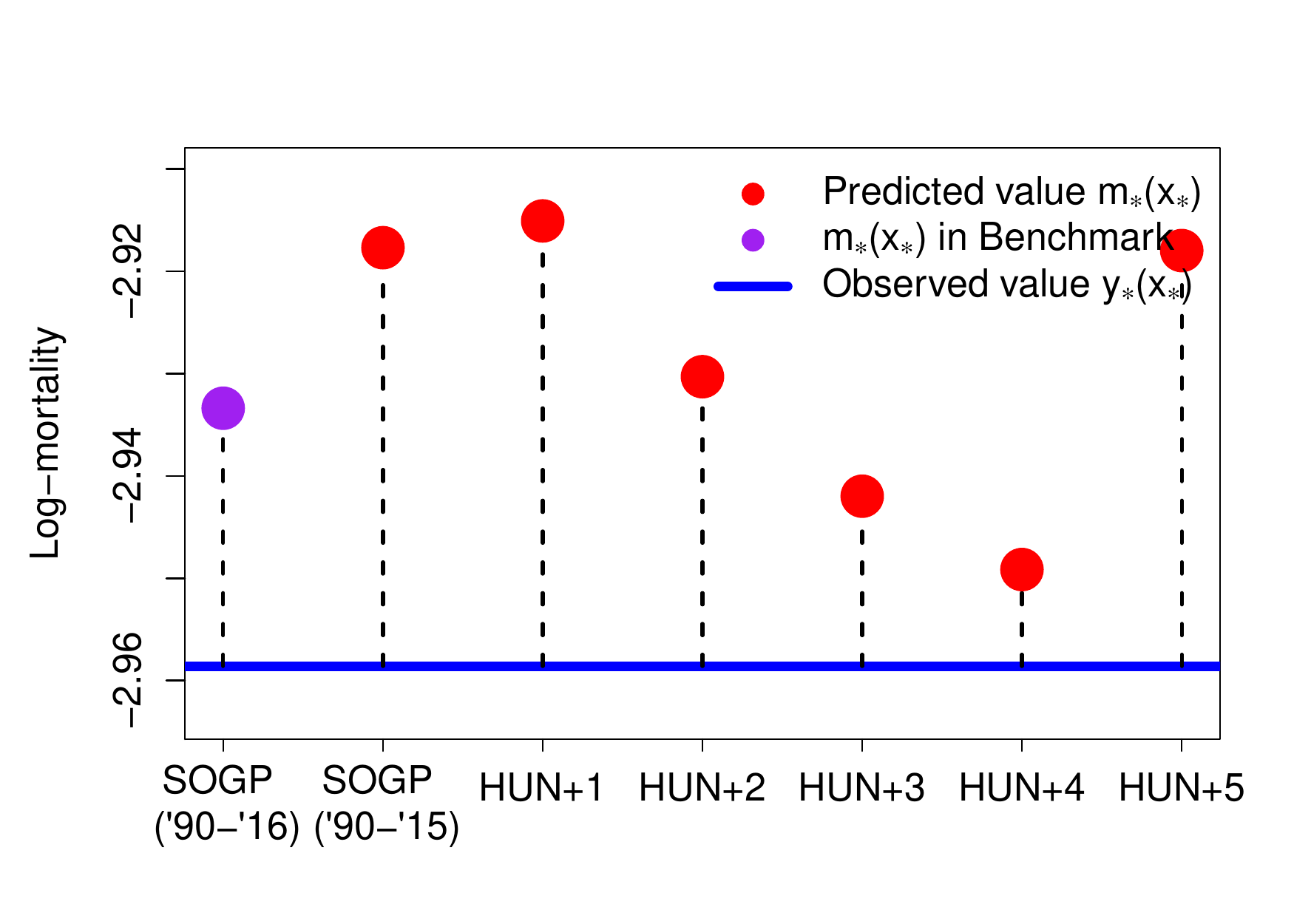}}} &
    {{\includegraphics[width=0.31\textwidth,trim=0in 0.1in 0in 0.1in]{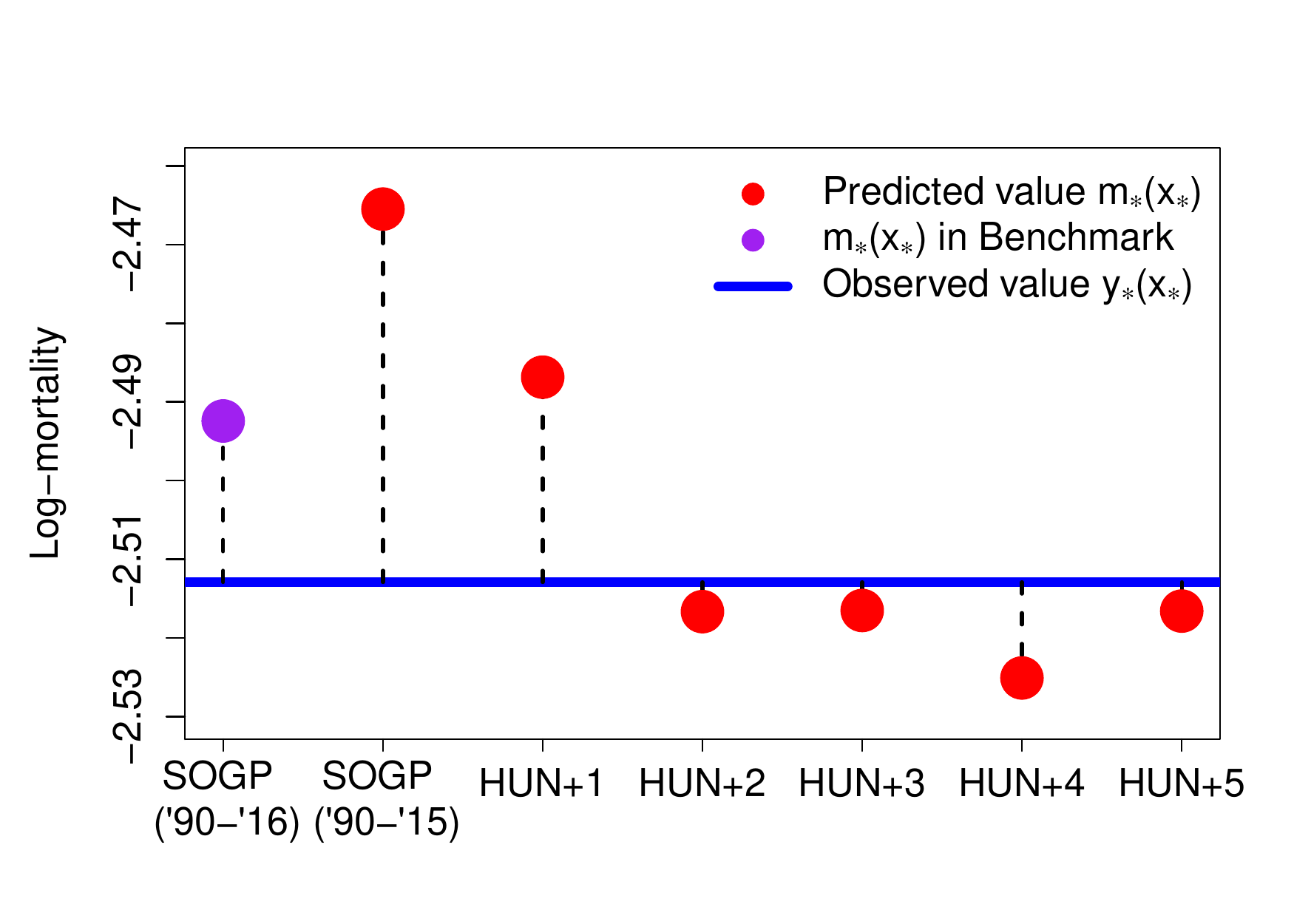}}} &
    {{\includegraphics[width=0.31\textwidth,trim=0in 0.1in 0in 0.1in]{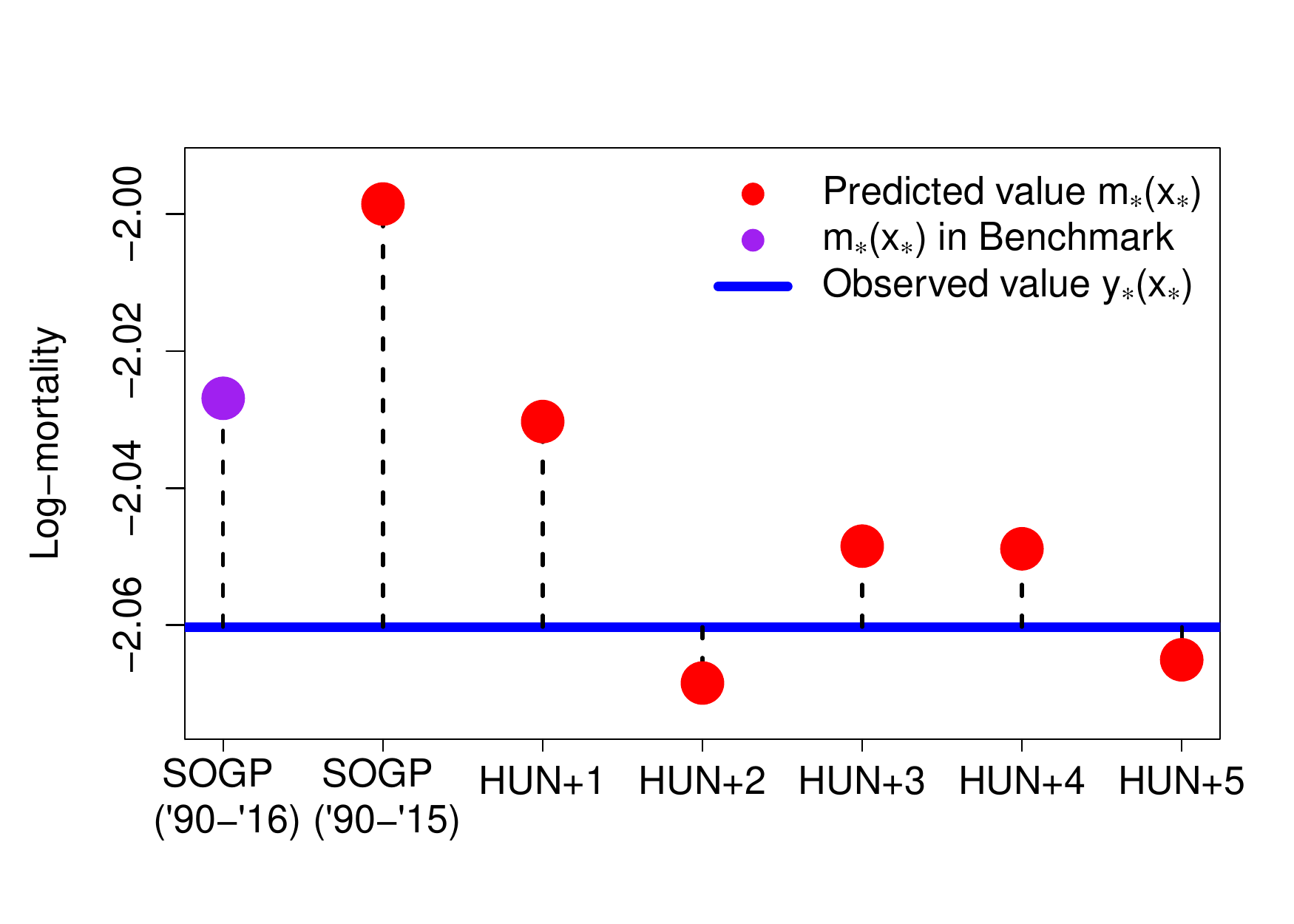}}} \\
    Age 74 & Age 79 & Age 84
    \end{tabular}
    \caption{Comparison of prediction accuracy for 2016 log-mortality of Hungarian Males between different ICM-MOGP models with ``notched'' setup. Top row: Standard deviation of $f(x_*)$; bottom row: distance between predicted mean $m_*(x_*)$ and the observed value $y_*(x_*)$.
    \label{fig:latest-data}}
\end{figure}

Moving beyond two populations,  in Figure~\ref{fig:latest-data}, we illustrate predictive gains for Hungarian Males due to incorporating most recent foreign data. This complements Table~\ref{tbl:manyPops-quality} that considered an isotropic dataset, with a notched setup instead. Our benchmark is a Hungary Males SOGP model fitted on 1990--2016. We then drop 2016 Hungary observations, but augment with 1990--2016 data from \emph{other} countries and perform 1-year-out extrapolation to forecast 2016 Hungary mortality. These models are labeled as `HUN+1', `HUN+2', etc., to indicate the number of foreign populations considered.  The top panels of Figure~\ref{fig:latest-data} visualize increasing forecast credibility, namely lower $s_*(x_*)$, for Hungary as more and more correlated data is added. In fact we see that a MOGP 1-year-out prediction with 3+ populations is \emph{more} credible than direct smoothing of realized 2016 Hungary experience. As expected, we observe that credibility gains flatten out as $L$ continues to grow and available information is saturated. The bottom panels of Figure~\ref{fig:latest-data} display the prediction errors $m_*(x_*)-y_*(x_*)$ relative to realized 2016 Hungary experience. Again, we see that higher $L$ tends to
generate less bias in prediction, confirm the earlier SMAPE analysis for isotropic case studies. Due to the strong observation noise the pattern for a specific cell $x_*$ can be erratic, although in nearly all cases, MOGP easily beats out the plain SOGP. To conclude, borrowing latest information from nearby highly-correlated populations is essentially as good as having the latest domestic data, and is significantly better than just using the available domestic data.

\emph{Remark:} In Figure~\ref{fig:latest-data} (and earlier in Table~\ref{tbl:manyPops-quality}) we add populations based on their correlation to the target population, i.e.~we pool through estimated $\theta_{l_1,l_2}$. In Appendix~\ref{sec:cluster-trend} we discuss a simpler alternative based on comparing mean functions that capture historical mortality trends and then running a hierarchical clustering method.

\section{Further Features of Multi-Output GP Models}\label{sec:features}

\subsection{Improved Hyperparameter Estimation}\label{sec:better-hyper}

\begin{figure}[ht]
\centering
    \subfloat[Length-scale in the Age dimension]{{\includegraphics[width=7.8cm,trim=0.2in 0in 0.2in 0.2in]{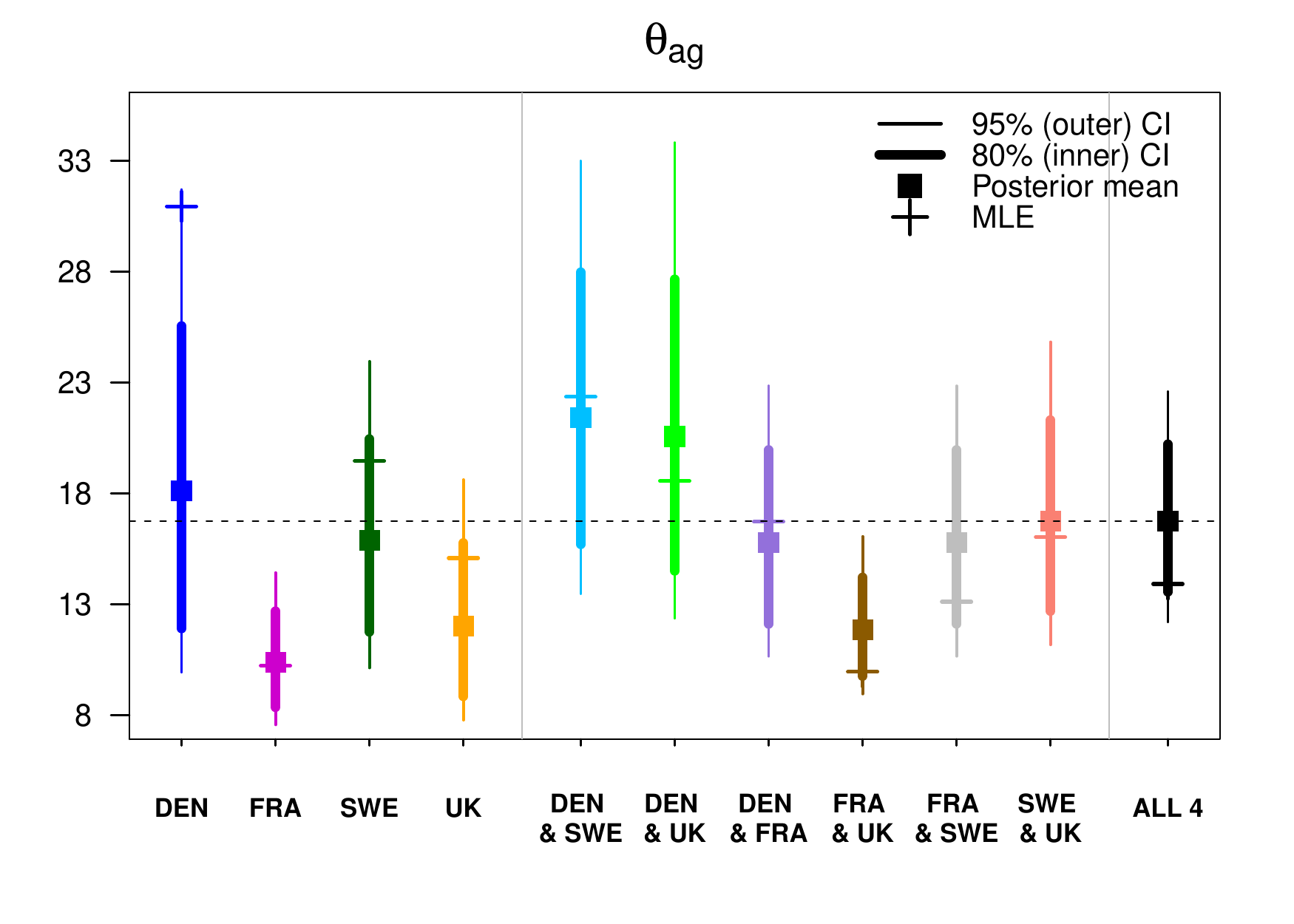}}}
    \qquad
    \subfloat[Length-scale in the Year dimension]{{\includegraphics[width=7.8cm,trim=0.2in 0in 0.2in 0.2in]{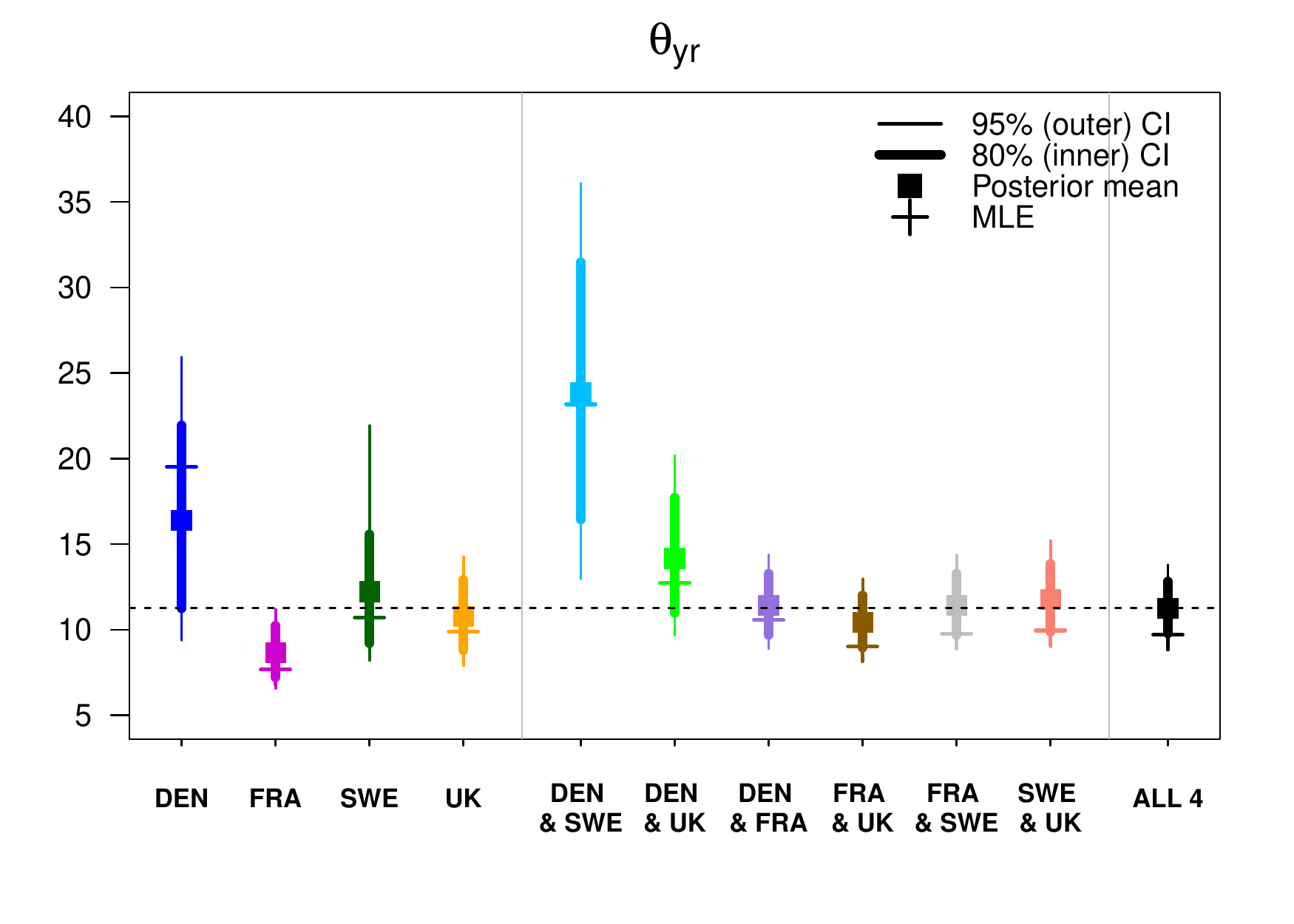}}}

    \caption{\code{Stan} MCMC posteriors of the lengthscales  $\theta$ for Age and Year across populations and joint models with different groupings. The $+$'s indicate the respective MLE estimates from a \code{kergp} model. The dashed lines indicate the MCMC MAP estimate from the 4-population Full MOGP model. }
    \label{fig:lscale-dis}
\end{figure}

To illustrate the commonality in  mortality experience in related populations we perform Bayesian GP on four developed Western European countries: Sweden, Denmark, France, and UK.  Figure \ref{fig:lscale-dis} shows the inferences of the lengthscales for Age and Year along with MLE estimations when fitting SOGP models for each population versus jointly modeling them as groups of two, or jointly as all 4 together. 
The figures visualize how joint GP models produce tighter hyperparameter posteriors. For example, the posterior mean of $\theta_{ag}$ in Denmark is relatively large and its credible bands are wide compared to the other three countries (Figure \ref{fig:lscale-dis}.a). However, once we pair Denmark with either Sweden, UK, or France (Figure \ref{fig:lscale-dis}.a ---light blue, light green, and purple CIs respectively), the credible bands of $\theta_{ag}$ become narrower and in the more reasonable range of $\theta_{ag} \in [15,30]$. This effect is even further amplified when taking all 4 countries together. The underlying concept is that
 the more populations are added into the model, the closer we get at discovering the ``universal'' representation of mortality pattern. In Figure \ref{fig:lscale-dis}, the 4-population MAP estimates of the lengthscales (dashed horizontal lines) intersect with a majority of CIs suggesting that there is indeed a common covariance structure which is gradually revealed as we increase the training dataset. We also remark that the MLE estimates fall within the 95\% posterior credible intervals from the \texttt{Stan} model indicating that Bayesian inference works properly.

This also highlights the ability of joint models to {better estimate the hyperparameters} by utilizing multiple data sets. 
It is known that GPs might have difficulties in estimating lengthscales, for example due to the likelihood function \eqref{eq:log-like} being highly multi-modal, or conversely very flat around its maxima. Providing more data is one remedy. As discussed in \cite{HLZ20} some SOGP will over- or under-smooth data while pooling data across multiple populations achieves shrinkage towards the global hyperparameter mean and provides a better fit.

\subsection{Coherent Mortality Forecasts}
\label{coherence}

\begin{figure}[!ht]
    \centering
    \captionsetup{justification=centering}
    \subfloat[Male/Female difference in log-mortality in Denmark using individual (SOGP) models]{{\includegraphics[width=6.5cm]{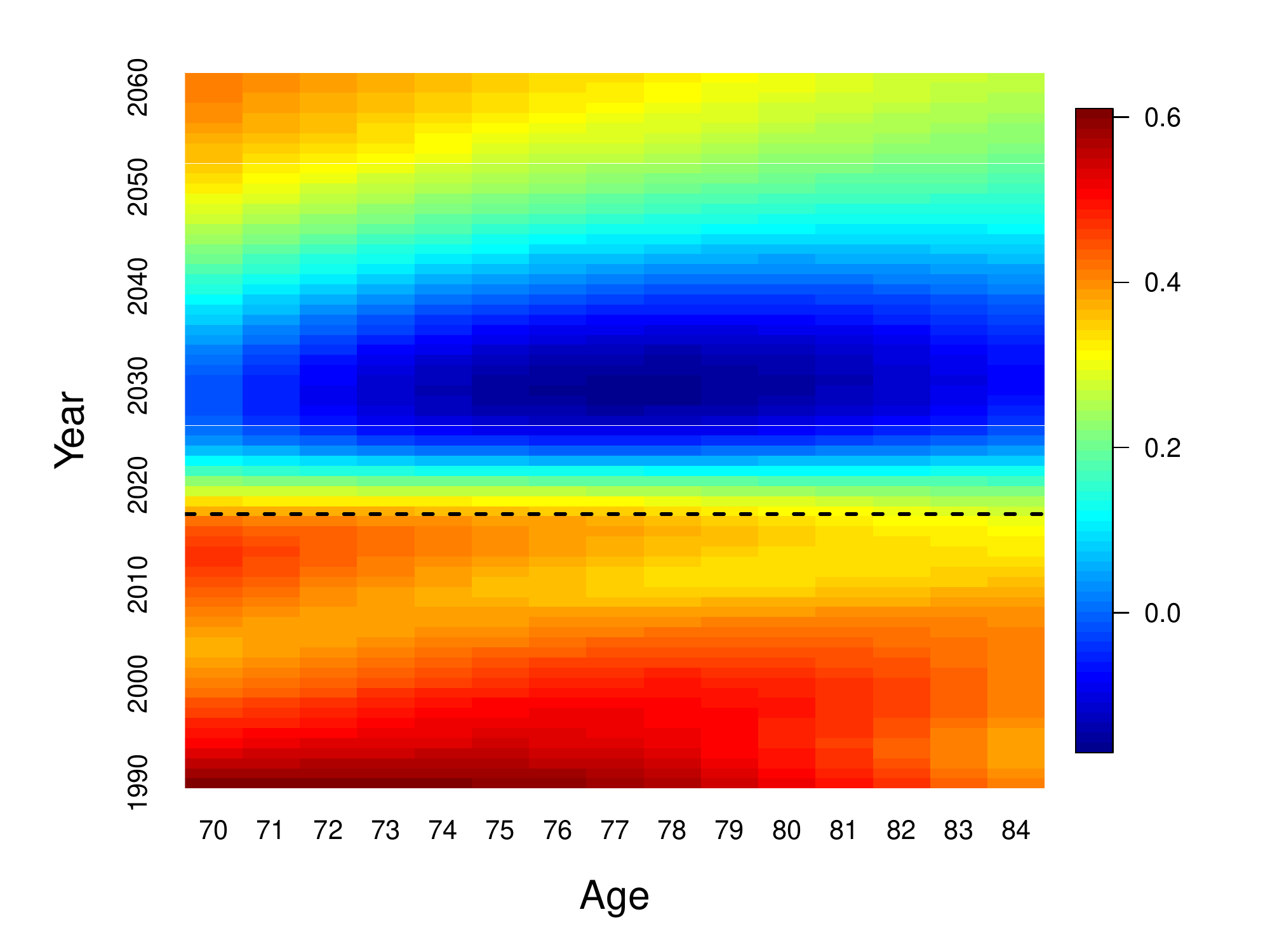}}}
    \qquad
    \subfloat[Male/Female difference in log-mortality in Denmark using a joint (Full-rank MOGP) model]{{\includegraphics[width=6.5cm]{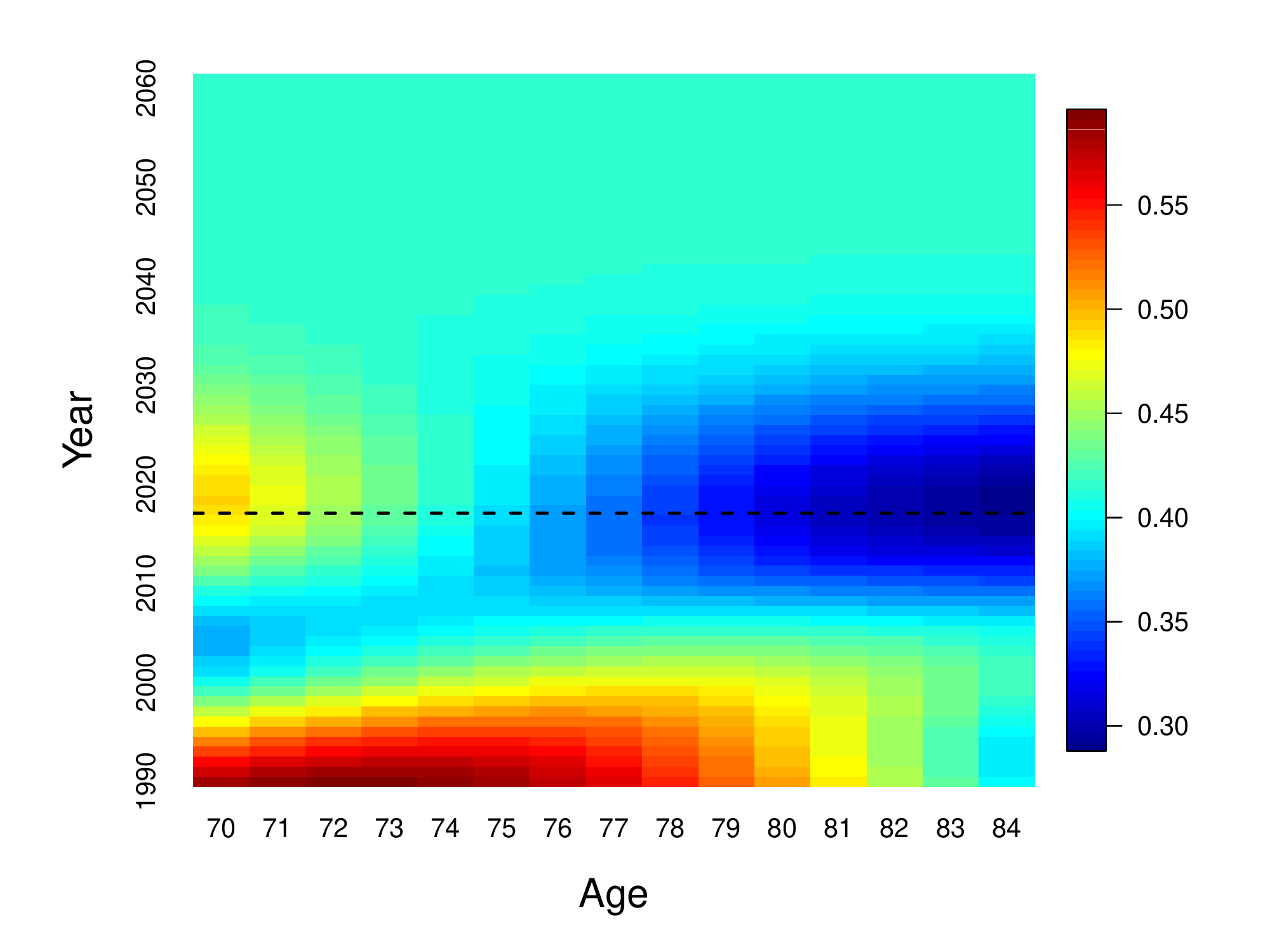}}}
    \caption{Forecasted mean difference between Danish Male and Female mortality over 1990 to 2060.   Training set was  Ages 70--84 and Years 1990--2016 (edge of training set indicated by the dashed lines).
    \label{fig:mort-convergence-gender}}
\end{figure}

Fitting GP models for individual populations tends to generate divergent long-term forecasts that are inconsistent with historical patterns. Multi-output GP models do not have this limitation and maintain the historical characteristics observed in the data into the future. Namely, in MOGP models, the long-term forecast is driven by the prior of $f$, and specifically by the mean function $m(\cdot)$. Thus, the relative differences in mortality between populations are controlled through the choice of $m(\cdot)$, so that different ways of achieving coherence are transparent to the modeler. For the linear mean function in \eqref{eq:mean-multi}, the population coefficients $\beta_{pop,l}$ serve this purpose and represent the long-term spread between same-Age log-mortality rates.
To illustrate the above, consider mortality differences due to gender. Women outlive men by 7 years on average in developed countries \citep{UnitedNation2011}. Modeling mortality for each gender separately fails to take into account this interdependence and tends to result in divergent and implausible long-run forecasts even if the same fitting procedure is applied.
%
%
%
The heatmaps in  Figure~\ref{fig:mort-convergence-gender} display the projected Male-Female differences in log-mortality for Denmark; single-population models on the left imply that as early as 2030, Males will have lower mortality than females. In contrast, the MOGP forecast in the right panel is coherent: Females are projected to maintain higher longevity and historical patterns slow dissipate over time to the long-term gap of $\beta_{pop,2}-\beta_{pop,1} = 41.5\%$ between same-age Male and Female mortality.

\begin{figure}[ht]
    \centering
    \subfloat[Predicted log-mortality rate]{{\includegraphics[width=0.47\textwidth]{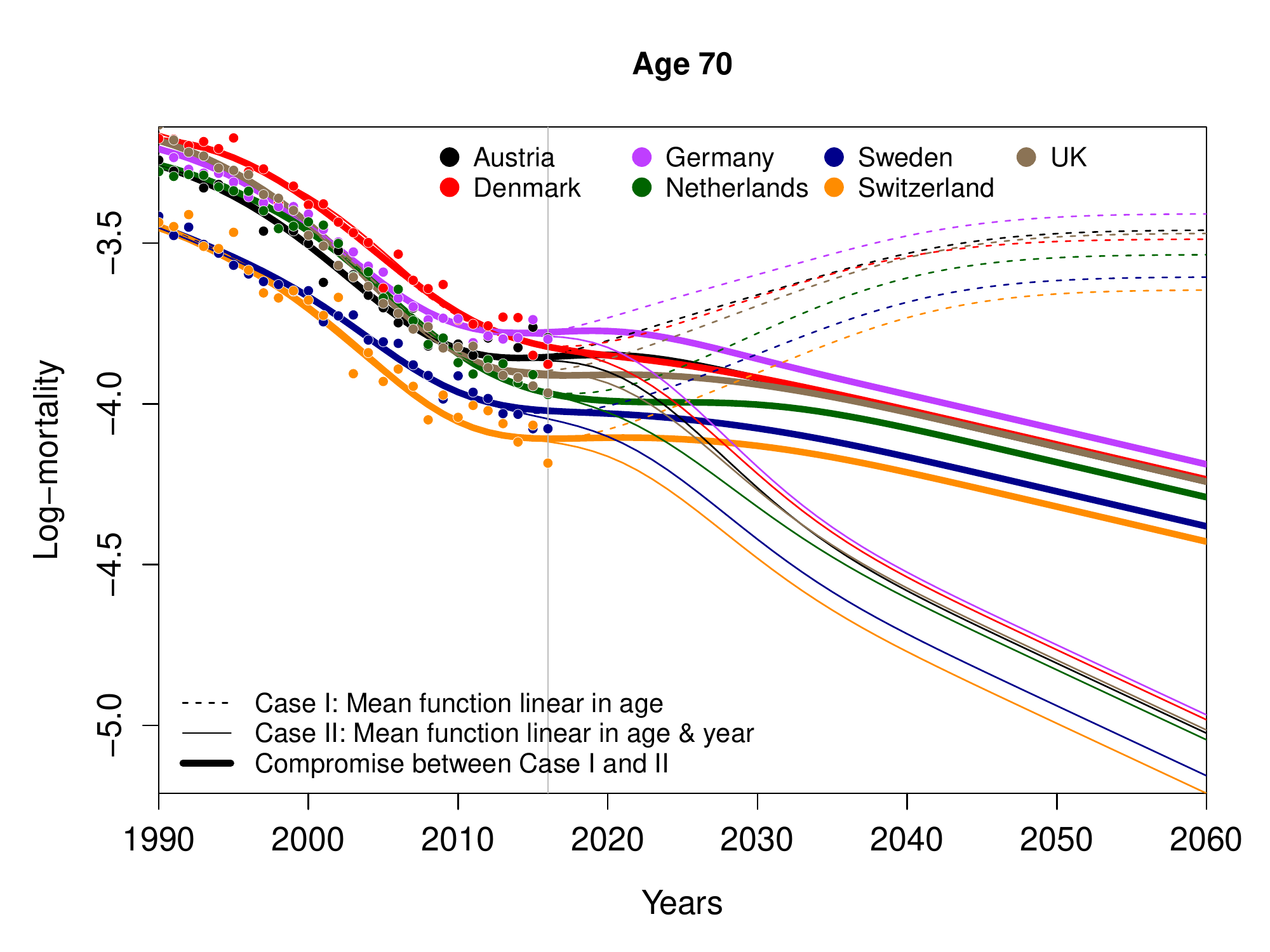}}}
    \qquad
    \subfloat[Annual mortality improvement factors]{{\includegraphics[width=0.47\textwidth]{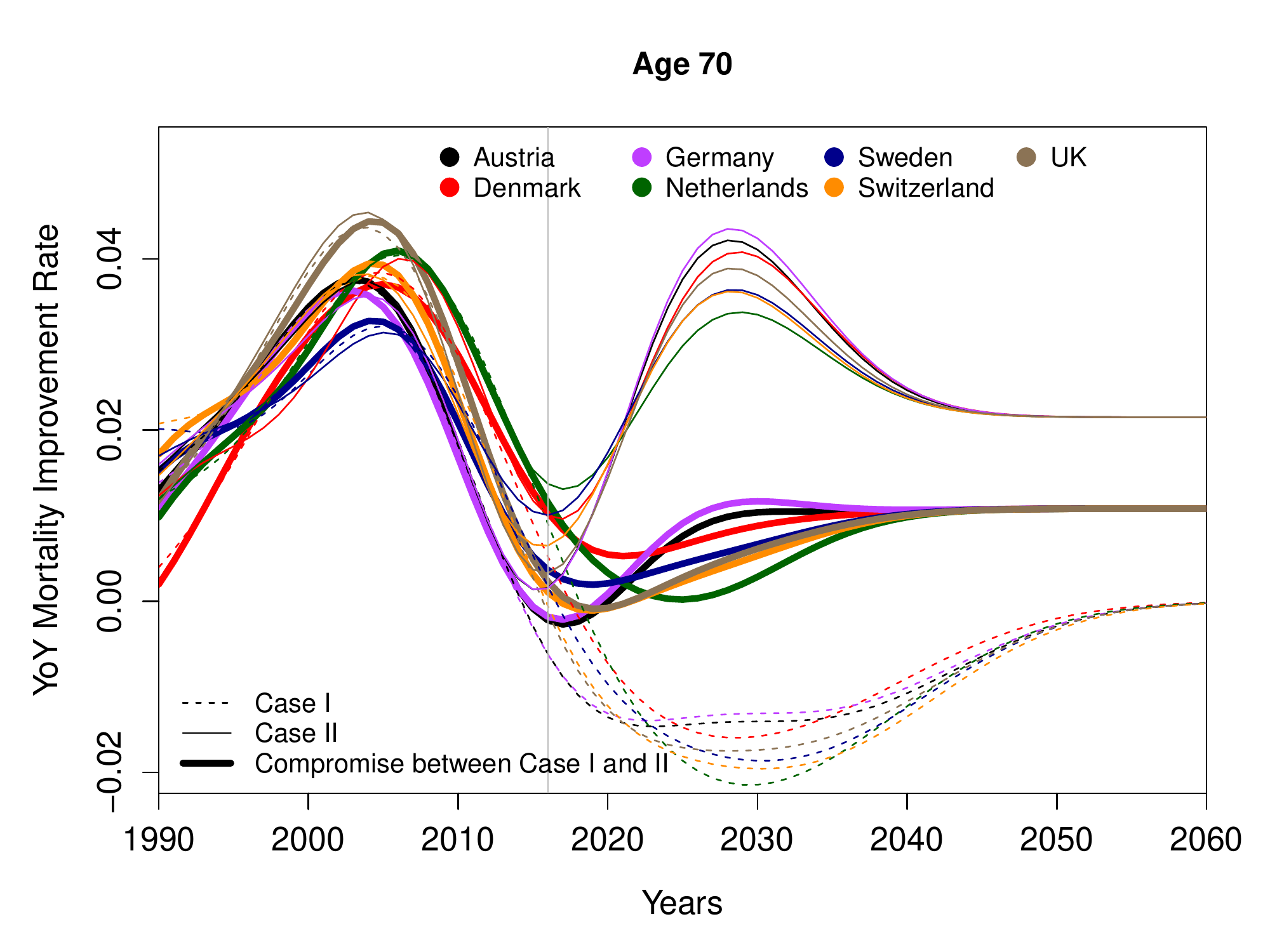}}}
    \caption{Long-term mortality forecasting over years 1990 to 2060.   The models are based on ICM kernel with $Q=3$ and are trained using Ages 70--84 and Years 1990--2016 over 7 Male populations: Austria, Denmark, Germany, Netherlands, Sweden, Switzerland and UK. Dashed lines indicate the boundary between in-sample and out-of-sample forecasts.
    \label{fig:mort-convergence}}
\end{figure}

 Figure \ref{fig:mort-convergence} shows the log-mortality and annual mortality improvement rates for Males aged 70 across seven European populations (indicated by colors), in the period from 1990 to 2060. We build a 7-population MOGP based on three different scenarios for $m(x)$: 
\begin{enumerate}
    \item Zero long-term mortality improvement, captured by the linear mean function $m(x^n)=\beta_0+\beta_1^{ag}x^n_{ag}+\sum_{l=2}^{L}  \beta_{pop,l} \big(x^n_{pop,l}\big)$ (dashed lines). All mortality improvement factors converge to zero (right panel) and the long-run mortality differences are summarized by the $\beta_{pop,l}$ coefficients.

    \item Long-term mortality improvement based on a historical pattern (thin solid curves). This is encapsulated via $ m(x^n)=\beta_0+\beta_1^{ag}x^n_{ag}+\beta_1^{yr}x^n_{yr}+ \sum_{l=2}^{L}  \beta_{pop,l} x^n_{pop,l}$. In the long-run $\partial m^{GP}_{back}(.;yr) \to \beta_1^{yr}$ (about 2\% annual); again $\beta_{pop,l}$ determine the long-run relative difference in longevity of different populations.

    \item Long-term mortality improvement based on expert judgement (thick solid lines). We again use $m(x^n)=\beta_0+\beta_1^{ag}x^n_{ag}+\beta_1^{yr}x^n_{yr}+ \sum_{l=2}^{L}  \beta_{pop,l} \big(x^n_{pop,l}\big)$, but this time the $\beta_1^{yr}$ coefficient is picked by the modeler and for illustrative purposes fixed at 1\% to reflect recent slowdown in global MI. Since it is not possible to fully extrapolate the future longevity trends from past data, it is appropriate to use expert opinions about future mortality \citep{Booth2008}.
\end{enumerate}

We observe that the choice of $m(\cdot)$ has minimal impact on in-sample forecasts that are largely driven by the training data covering 1990--2016. On the other hand, the long-term levels of mortality improvement are \emph{completely} driven by $m(\cdot)$. Finally, for short-term extrapolation (roughly 2016--2025 in the Figure, reflecting the fitted Year lengthscale $\theta_{yr} \simeq 10$) the forecasts blend information from the training set and from $m(\cdot)$. Note that in this example some of the individual mortality curves may cross, i.e.~the relative order of longevity in different populations may change over time (such as Denmark surpassing Germany's longevity) due to higher recent improvement rates. Nevertheless, we see a very strong coherence so that mortality rates across populations all move roughly in unison over time, matching our intuition about the persistent commonality of their future mortality experiences.

\section{Conclusion}\label{sec:conclusion}

We have developed and investigated stochastic multi-population mortality models based on multi-output Gaussian process regression. In our approach, cross-population dependence is captured via spatial correlation that overlays the Age-Year structure. This yields a unified approach for any number of populations $L$; moreover the proposed coregionalization kernels allow to leverage the Kronecker structure and incorporate dimension reduction for the underlying cross-population factors. Our analysis of HMD data suggests that the MOGP approach is well-suited to selectively fuse mortality experience from similar datasets, where similarity can be interpreted through the spatial GP correlations $r_{l_1,l_2}$.  On the one hand, we find that pooling disparate populations can be counter-productive (since MOGP relies on the assumption of homogenous Age-Year covariance pattern); on the other hand, pooling can indeed yield significant improvement in predictive accuracy, especially in smaller populations with low credibility.

Looking ahead, it would be worthwhile to investigate large-scale models, e.g.~based on the full HMD database of 40 countries and 2 genders. This requires additional modeling infrastructure as the presented approach becomes computationally expensive for $L \gg 10$ populations (more than $N \gg 5000$ total cells). There is currently a very active and ongoing progress on large-scale GP models especially for gridded data like in HMD, see e.g.~\cite{flaxman2015fast}. A different avenue of future research would be to systematically explore the best spatial covariance structures, as encapsulated by the kernel function $\tilde{C}(x, x')$. In this paper we focused on only using the squared-exponential kernel and standard Age- and Year-effects. It is feasible to consider further dependence formats, e.g.~birth Cohort effect, and other kernel families, such as the Matern \cite{LUDKOVSKI2016GAUSSIAN}. A third direction would be to revisit the observation variance assumption via GLM (generalized linear model) GP formulations.

\appendix

\section{Clustering by Mortality Trends}\label{sec:cluster-trend}

In Section \ref{cluster} we constructed multi-population MOGPs by first generating a large set of  2-population models and then utilizing the respective correlations $r_{l_1,l_2}$ to select the datasets that are most correlated to the target population. We also investigated a simpler alternative for deciding which populations to pool based on the similarity in their mortality trends. This approach does not require construction of any preliminary MOGPs and instead only looks at the GP mean function $m(\cdot)$. Namely, we first estimate the shape of mortality for each population $l$ via a linear mean function: $m_l(x)=\beta_{0,l} +\beta_{1,l}^{ag}x_{ag}+\beta_{1,l}^{yr}x_{ag}$. 
We then  calculate the root integrated squared distance between $m_{l_1}(\cdot), m_{l_2}(\cdot)$ based on the given test set $x_{ag} \in \{50,\ldots,84\}$ and $x_{yr} \in \{1990,...,2012\}$:
    \begin{align}\label{eq:D-metric}
        D_{l_1,l_2} & 
        & =\sqrt{\int_{1990}^{2012}\int_{50}^{84}\bigg[(\beta_{0,l_1}-\beta_{0,l_2})+(\beta_{1,l_1}^{ag}-\beta_{1,l_2}^{ag})x_{ag}
        +(\beta_{1,l_1}^{yr}-\beta_{1,l_2}^{yr})x_{yr}\bigg]^2dx_{ag}dx_{yr}}.
    \end{align}

\begin{figure}[!h]
        \centering
        \subfloat[Complete linkage]{{\includegraphics[width=6.5cm]{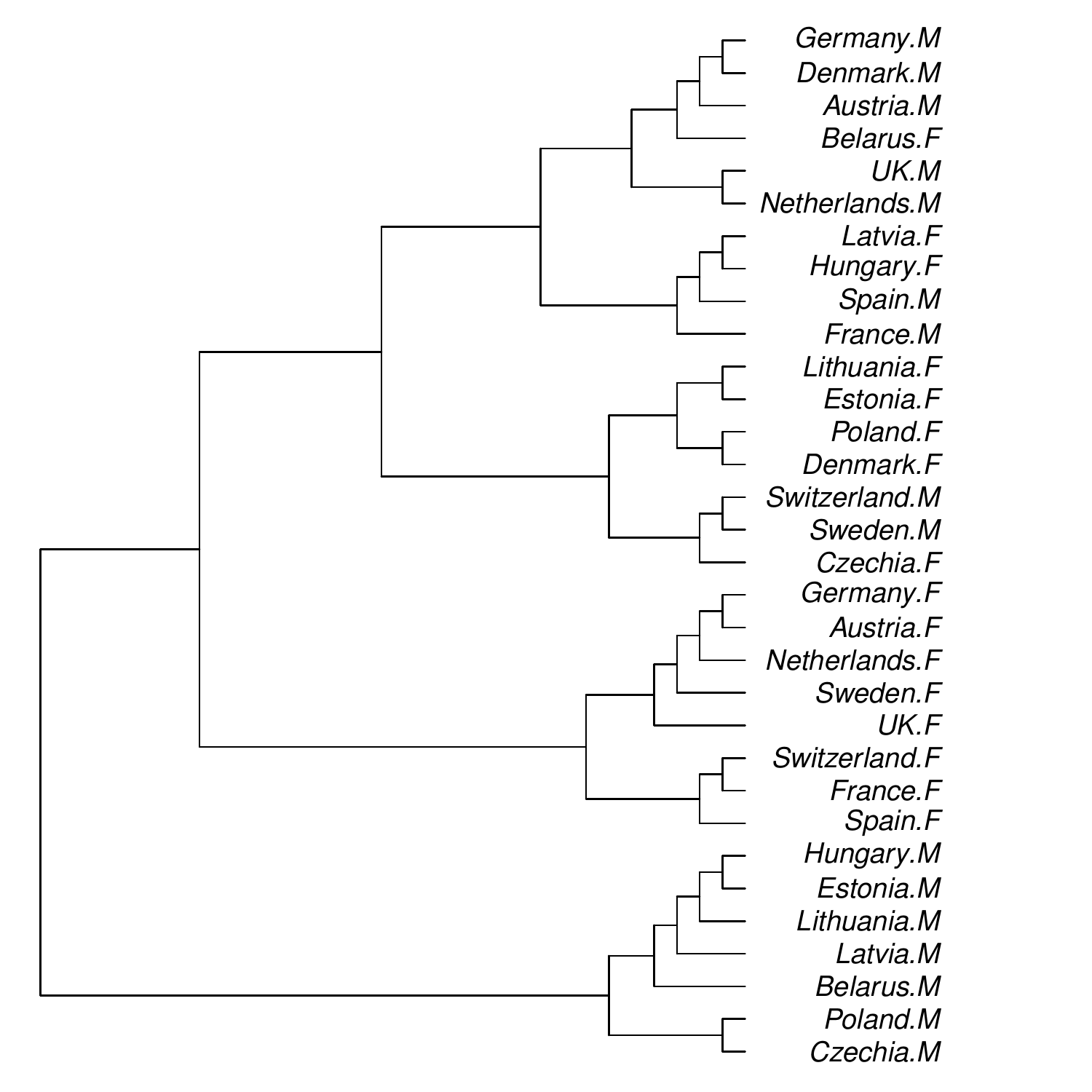}}}
        \subfloat[Single linkage]{{\includegraphics[width=6.5cm]{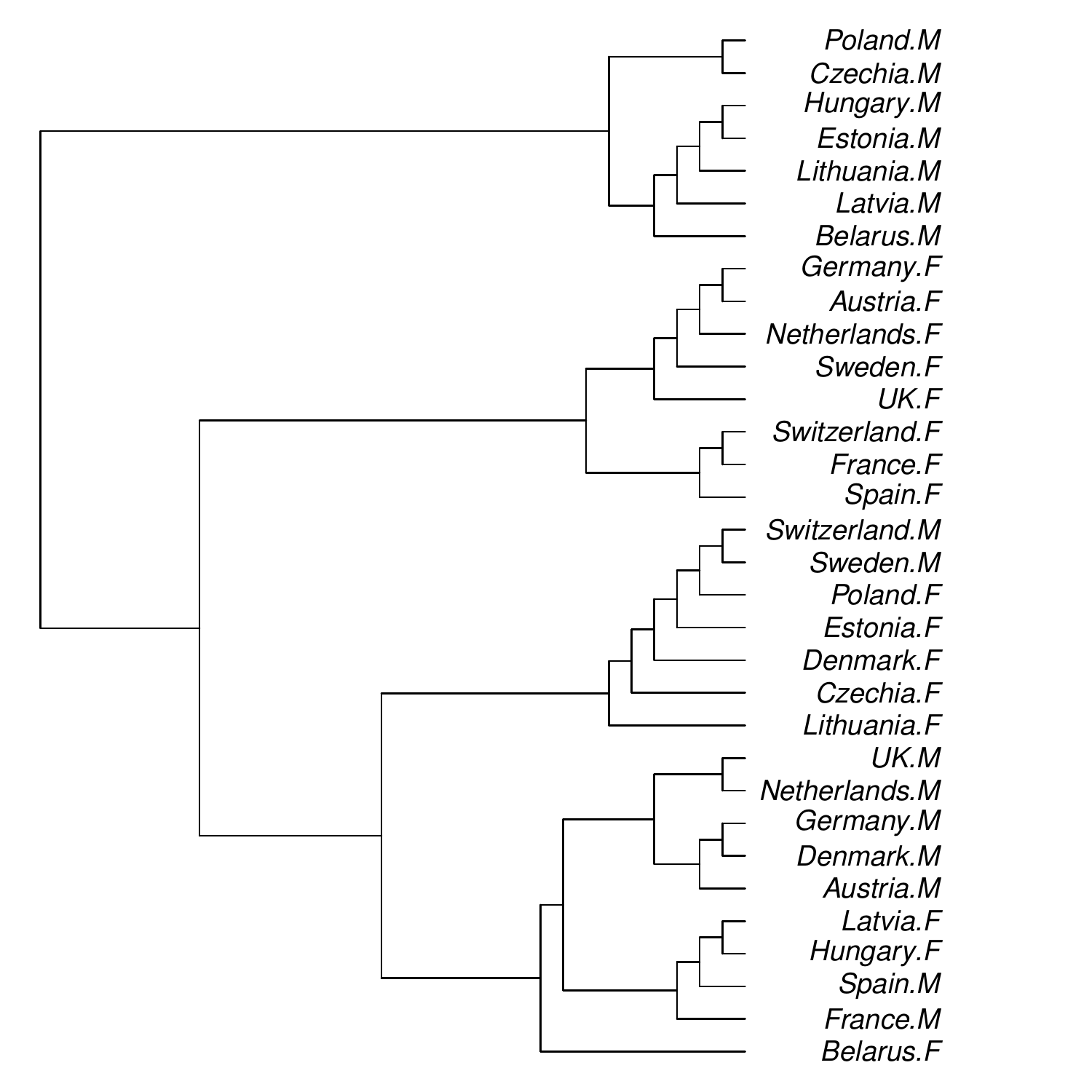}}}
        \caption{Dendrograms from hierarchical clustering of 32 HMD populations using the $D$-metric in \eqref{eq:D-metric}.}
        \label{fig:cluster}
    \end{figure}

    The above metric is employed within a hierarchical clustering method based on a specified dissimilarity measure. Figure~\ref{fig:cluster} displays the dendrograms extracted from hierarchical clustering of 32 populations (16 countries $\times$ \{Male, Female\}) via two different measures of dissimilarity: single linkage and complete linkage. We note that the resulting clusters naturally tend to separate males and females, reflecting the latter's lower mortality. We also observe a strong geographic influence, so that neighboring countries with similar demographics are indeed clustered together. The dendrogram could be used agglomeratively to build a MOGP. For example, using Hungary as the target population, Figure~\ref{fig:cluster} suggests first adding Estonia, then Lithuania, Latvia, etc. We find that this method is often as efficient as clustering by cross-population correlation, although going up the linkages often calls for fusing of two clusters, i.e.~it is not designed for increasing $L$ 1-by-1.

\section{Fitted MOGP hyperparameters across Full-Rank and ICM kernels and MLE/Bayesian methods for the DEN-FRA-SWE-GBR Case Study}

\begin{table*}[!th]
\centering
\caption{Hyper-parameter estimates based on maximum likelihood (\code{kergp}) and maximum a posteriori probability (\code{Stan}), along with MCMC summary statistics using a joint mortality model across four countries: Denmark, Sweden, France, and UK. Training set contains Males aged 70--84 during Years 1990--2012.  Denmark used as baseline population in the mean function.}
\label{tbl:gpB-joint4}
\begin{tabular}{lrrrr}\toprule
\multirow{2}{*}{Parameters} & \multirow{2}{*}{\begin{tabular}[c]{@{}c@{}}\code{kergp}\\ MLE\end{tabular}} & \multicolumn{3}{c|}{\texttt{Stan}}                \\ \cline{3-5}
&   & MAP & MCMC Mean & MCMC 95\% Posterior CI \\ \hline
$\beta_0$       & $-$10.4610    & $-$10.0220   & $-$10.5337      & ($-$12.0847, $-$9.1274)         \\
$\beta_1^{ag}$  &  0.0996    & 0.0958     & 0.0967       & (0.0847, 0.1085)          \\
$\beta_{FRA}$  & $-$0.0922   & $-$0.0685    & 0.1239      & ($-$0.2438,  0.5827) \\
$\beta_{SWE}$ & $-$0.1076  & $-$0.0592    & $-$0.0060      & ($-$0.3596,  0.3844) \\
$\beta_{GBR}$ & 0.0062  & 0.000   & 0.1122     & ($-$0.2252,  0.4961) \\
$\theta_{ag}$   & 13.3849  &  12.1915   & 17.4166 & (12.0294, 24.0641)  \\
$\theta_{yr}$   & 8.8549    & 9.2694  & 11.3858   & (8.2536, 13.3009)  \\
$\theta_{DEN,~FRA}$ & 0.1956  &  0.3773     & 0.8269   & (0.1544, 2.9089)           \\
$\theta_{DEN,~SWE}$ & 0.1804  &   0.2725 & 0.5094 & (0.0889, 1.8891)           \\
$\theta_{DEN,~GBR}$ & 0.0772  &  0.0799      & 0.1579    & (0.0286, 0.5473)           \\
$\theta_{FRA,~SWE}$ & 0.1030 &  0.1943      &  0.3658  & (0.0797, 1.0949)           \\
$\theta_{FRA,~GBR}$ & 0.1095  &  0.1445      & 0.1439   & (0.0383, 0.3917)           \\
$\theta_{SWE,~GBR}$ & 0.1681  &  0.1801      &  0.6132  & (0.0530, 2.6660)           \\
$\eta^2$        &  0.0395  & 0.0392 & 0.0684 & (0.0289, 0.1520)\\
$\sigma^2_{DEN}$      & \num{1.516e-03}   & \num{1.514e-03}  &  \num{1.528e-03 } &  (\num{1.315e-03}, \num{1.772e-03})       \\
$\sigma^2_{FRA}$      & \num{3.394e-04}   & \num{3.371e-04}  & \num{3.459e-04}  & (\num{2.956e-04}, \num{4.045e-04})\\
$\sigma^2_{SWE}$      & \num{8.022e-04}   & \num{8.007e-04} & \num{8.226e-04}  &  (\num{7.033e-04}, \num{9.640e-04})        \\
$\sigma^2_{GBR}$      & \num{6.887e-04}   & \num{6.849e-04} & \num{7.001e-04}   & (\num{5.985e-04}, \num{8.165e-04})         \\\bottomrule
\end{tabular}

\end{table*}

\begin{table}[!bh]
\centering
\caption{MOGP with rank $Q=3$ ICM kernel for Males in Denmark, France, Sweden, and UK. The  training dataset contains Ages 70--84 and Years 1990--2012.}
\begin{tabular}{ccclccccc} \toprule
\multicolumn{3}{c}{Mean function}          &  & Lengthscales & Cross-covariance & $a_{l,1}$ & $a_{l,2}$ & $a_{l,3}$ \\ \cline{1-3} \cline{5-9}
$\beta_0$  & $\beta_{ag}$  & $\beta_{FRA}$ &  & $\theta_{ag}$     & DEN: $l=1$       & 0.0435    & 0.1687    & 0.1068    \\
$-$11.4073 & 0.1120        & 0.0410        &  & 15.4199       & FRA: $l=2$       & 0.2199    & 0.1619    & 0.0491    \\
$-$        & $\beta_{SWE}$ & $\beta_{GBR}$ &  & $\theta_{yr}$    & SWE: $l=3$       & 0.1552    & 0.0562    & 0.1683    \\
$-$        & $-$0.0348     & 0.0520        &  & 13.4019     & GBR: $l=4$       & 0.1451    & 0.1636    & 0.1640    \\  \bottomrule
\end{tabular}
\end{table}

\textbf{Cross-population correlation matrices.} In the full-rank kernel \eqref{eqn:kernelMulti}, the correlation coefficient is $r_{l_1,l_2}:=\exp{(-\theta_{l_1,l_2})}$. In the ICM model the cross-covariance is $B = A A^T$ with the diagonals $B_{l,l}$ interpreted as the process variance of $f_l$, from which we can similarly extract $r_{l_1,l_2}$'s. Using \{Denmark, France, Sweden, UK\} $\equiv \{1,2,3,4\}$ we find that
\begin{align*}
\begin{bmatrix}
        r_{21} &  &    \\
        r_{31} & r_{32} &   \\
        r_{41} & r_{42} & r_{43} \\
    \end{bmatrix}
    & = \begin{bmatrix}
        0.82 &   &  \\
        0.83 & 0.90 &  \\
        0.93 & 0.90 & 0.84  \\
    \end{bmatrix} \text{for full rank; and }\\  &= \begin{bmatrix}
        0.75 &   &  \\
        0.73 & 0.84 &  \\
        0.96 & 0.98 & 0.89  \\
    \end{bmatrix} \text{ for ICM}.
\end{align*}

\bibliographystyle{plainnat}

\bibliography{biblio}

\end{document}